\documentclass[sigconf]{acmart}
\usepackage{bbm}
\usepackage[ruled,linesnumbered]{algorithm2e}
\usepackage{multirow}
\usepackage{graphicx}
\usepackage{subfig}
\usepackage{enumitem}
\usepackage{hyperref}
\usepackage{float}
\usepackage{stfloats}
\hypersetup{
    colorlinks=true,
    linkcolor=blue,
    filecolor=blue,      
    urlcolor=blue,
    citecolor=cyan,
}
\newcommand{\SEBot}{\textsc{SeBot}\xspace}

\newcommand{\mypara}[1]{{\noindent\textbf{#1}}}

\AtBeginDocument{%
  \providecommand\BibTeX{{%
    \normalfont B\kern-0.5em{\scshape i\kern-0.25em b}\kern-0.8em\TeX}}}
\copyrightyear{2024}
\acmYear{2024}
\setcopyright{acmlicensed}
\acmConference[KDD '24]{The 30th SIGKDD Conference on Knowledge Discovery and Data Mining}{August 25--29, 2024}{Spain}
\acmBooktitle{The 30th SIGKDD Conference on Knowledge Discovery and Data Mining (KDD '24), August 25--29, 2024, Spain}
\acmPrice{15.00}
\acmDOI{10.1145/xxxxxxx.xxxxxxx}
\acmISBN{978-1-4503-9416-1/23/04}

\begin{document}

\title{\SEBot: Structural Entropy Guided Multi-View Contrastive Learning for Social Bot Detection}

\author{Yingguang Yang}
\orcid{0000-0002-2473-6229}
\affiliation{%
  \institution{University of Science and Technology of China}
  \country{}
}
\email{dao@mail.ustc.edu.cn}

\author{Qi Wu}
\affiliation{%
  \institution{University of Science and Technology of China}
  \country{}
}
\email{qiwu4512@mail.ustc.edu.cn}

\author{Buyun He}
\affiliation{%
  \institution{University of Science and Technology of China}
  \country{}
}
\email{byhe@mail.ustc.edu.cn}

\author{Hao Peng}
\affiliation{%
 \institution{Beihang University}
 \country{}
 }
\email{penghao@buaa.edu.cn}
\authornote{Corresponding authors}

\author{Renyu Yang}
\affiliation{%
 \institution{Beihang University}
 \country{}
 }
\email{renyu.yang@buaa.edu.cn}

\author{Zhifeng Hao}
\affiliation{%
  \institution{Shantou University}
  \country{}
  }
\email{haozhifeng@stu.edu.cn}

\author{Yong Liao}
\affiliation{%
  \institution{University of Science and Technology of China}
  \country{}
}
\email{yliao@ustc.edu.cn}
\authornotemark[1]

\renewcommand{\shortauthors}{Yingguang Yang et at.}

\begin{abstract}
Recent advancements in social bot detection have been driven by the adoption of Graph Neural Networks. The social graph, constructed from social network interactions, contains benign and bot accounts that influence each other. However, previous graph-based detection methods that follow the transductive message-passing paradigm may not fully utilize hidden graph information and are vulnerable to adversarial bot behavior. The indiscriminate message passing between nodes from different categories and communities results in excessively homogeneous node representations, ultimately reducing the effectiveness of social bot detectors. In this paper, we propose \SEBot, a novel multi-view graph-based contrastive learning-enabled social bot detector. In particular, we use structural entropy as an uncertainty metric to optimize the entire graph's structure and subgraph-level granularity, revealing the implicitly existing hierarchical community structure. And we design an encoder to enable message passing beyond the homophily assumption, enhancing robustness to adversarial behaviors of social bots. Finally, we employ multi-view contrastive learning to maximize mutual information between different views and enhance the detection performance through multi-task learning.
Experimental results demonstrate that our approach significantly improves the performance of social bot detection compared with SOTA methods.
\end{abstract}




\maketitle
\section{Introduction}
\begin{figure}
    \centering
    \includegraphics[width=0.9\linewidth]{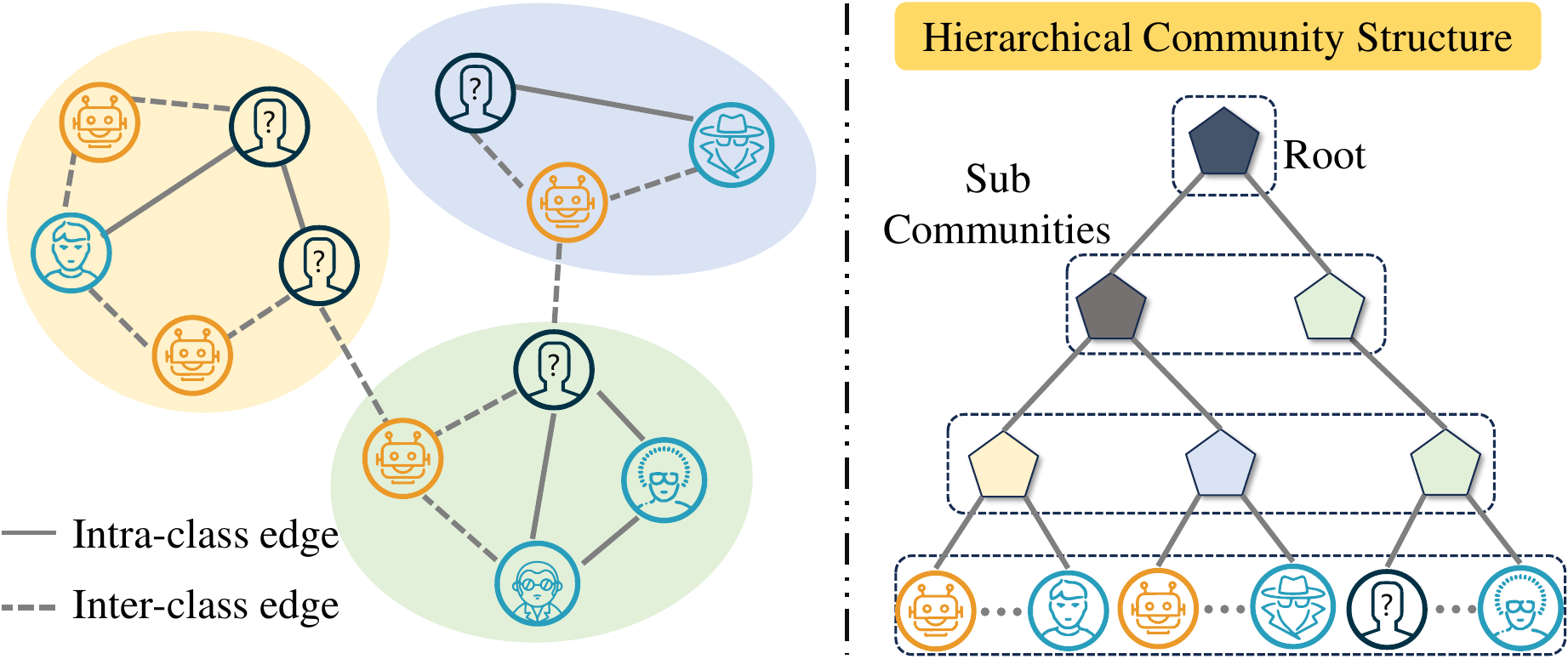}
    \caption{Illustration of community structure and inter-class interactions co-exists in social networks. The abstraction of the hierarchical community structure is presented in the form of an encoding tree.}
    \label{fig: intro}
\end{figure}
Social bots are automated controlled accounts that widely exist on social platforms such as Twitter and Weibo, in most cases, for malicious purposes such as spreading misinformation \cite{ferrara2017disinformation, wang2018era}, manipulating public opinion \cite{hamdi2022mining,weng2022public}, and influencing political elections \cite{deb2019perils,luceri2019red}. They will undoubtedly give rise to societal disharmony in social network environments. 

Conventional methods for social bots detection primarily focus on extracting discriminative features, ranging from user attributes, \cite{wu2021novel}, text features \cite{heidari2020deep} to structure features\cite{cao2015targetvue}, which can be then used to train classifiers in a supervised manner. Due to the continuous evolution of social bots~\cite{cresci2020decade}, including their ability to steal information from legitimate accounts and mimic normal account behaviors~\cite{le2022socialbots}, these traditional methods turn out to be ineffective in identifying the latest generation bots. A recent advancement in social bot detection is introducing graph neural networks ~\cite{ali2019detect} that treat accounts and the interactions in-between as nodes and edges, respectively. Multi-relational heterogeneous graphs can be established \cite{feng2022heterogeneity} and a Relation Graph Transformer (RGT) is responsible for aggregating information from neighbors. Such approaches consider the features of each account as interdependent ones and leverage the semantic information of relationships between accounts to generate semantically-richer representations.

While successful, there are two challenges facing the existing graph-based methods (illustrated in Figure \ref{fig: intro}).  \textit{i) How to fully exploit the hierarchical information hidden in the graph structure?} Unlike other semi-supervised node classification tasks, some accounts in social platforms tend to exhibit stronger correlations with others in the topological structure due to shared interests, events of interest, etc.~\cite{cresci2020decade}, indicating an intrinsic hierarchical community structure in the graph constructed by social bot detectors.  The existing graph-based social bot detection methods~\cite{feng2021botrgcn, feng2022heterogeneity, wu2023heterophily, yang2023rosgas, yang2023fedack,he2024dynamicity,peng2024unsupervised,zeng2024adversarial} primarily focused on aggregating node-level information on the original graph, neglecting  comprehensive utilization of high-order graph structural information. 
As a result, the generated representations only capture low-order graph information and lack high-order semantic information. \textit{ii) How to handle the adversarial behaviors of bots intentionally interacting with humans to evade detection?}  Existing bot detection methods rely on the assumption that bots and humans exhibit stronger connections with nodes of their category.  However, for social bots, especially in the case of advanced social bot control programs, gathering human-specific information from neighbors would be more advantageous for concealing their true identity. Therefore, bots intentionally tend to establish associations with human accounts to escape  detection~\cite{chavoshi2017temporal,des2022detecting}. 


To tackle these two challenges, we propose \SEBot, a novel \underline{S}tructural \underline{E}ntropy Guided Social \underline{Bot} Detection framework through graph contrastive learning enhanced classification with both intra-class and inter-class edges. Motivated by minimum entropy theory \cite{jaynes1980minimum}, indicating systems at the minimum level of uncertainties, and structural entropy \cite{li2016structural} further offers an effective measure of the information embedded in an arbitrary graph and structural diversity. We construct encoding trees for both the entire input graph and the subgraphs of each node by minimizing their structural entropy. We then use the optimal encoding trees to describe the hierarchical structural information of the graph and obtain cluster assignment matrixes for the nodes. 
Subsequently, node representations are obtained through structural entropy pooling and unpooling for node and graph classification. On the other hand, we designed an encoder that can adaptively make neighbors similar or differentiate them to handle the adversarial behavior of bots. Finally, we utilize the node representations generated by the three modules to calculate both cross-entropy loss and self-supervised contrastive learning loss. 

The contributions of this work can be summarized as follows:
\begin{itemize}  [leftmargin=*]
    \item We are the first to introduce structural entropy to capture semantic information at both subgraph and entire-graph levels in a self-supervised manner.
    \item We propose a self-supervised contrastive learning framework called \SEBot, which integrates node-node and subgraph-subgraph level contrastive learning tasks in a multi-task learning manner to capture high-order semantic information.
    \item Experiments on two real-world bot detection benchmark datasets demonstrate that our method outperforms current state-of-the-art social bot detection methods.
\end{itemize}

\section{Related Work}
\mypara{Graph-based Social Bot Detection}. Graph-based social bot detection has been of ultimate importance in modeling various interactions intrinsically existing in social networks. Previous methods~\cite{ali2019detect, feng2021botrgcn, feng2022heterogeneity,wu2023heterophily} have focused primarily on designing information aggregation strategies for better detection performance. ~\cite{ali2019detect} takes the first attempt to use graph convolutional neural networks (GCNs)~\cite{kipf2016semi} for detecting social bots. Typically, BotRGCN ~\cite{feng2021botrgcn} utilizes relational graph convolutional networks (RGCNs)~\cite{schlichtkrull2018modeling} to aggregate neighbor information from edges of different relations.  ~\cite{feng2022heterogeneity} proposed RGT, which utilizes a self-attention mechanism to adaptively aggregate information from neighbors in each relational view of the graph.  Although these methods have shown significant improvements compared to traditional feature engineering and text-based approaches~\cite{feng2022twibot}, they may not fully exploit the crucial semantic information concealed in the graph structure and the graph structure obtained from sampling social networks contain a significant amount of uncertainty and randomness.

\mypara{Graph Self-Supervised Learning}. Self-supervised learning has achieved great success in the fields of natural language processing ~\cite{yan2021consert} and computer vision ~\cite{kang2020contragan} without the need for prohibitively costly labeled data. Graph contrastive learning (GCL) is a typical paradigm of self-supervised learning on graphs, aiming at learning invariant representations between different graph views. For instance, DGI \cite{velickovic2019deep} utilizes a local-global mutual information maximization approach to obtain node representations. \cite{you2020graph} proposes a series of graph augmentation methods including node dropping, edge perturbation, attribute masking, and so on. \cite{zhu2021graph} propose an adaptive way for graph augmentation, which assigns adaptive probabilities to attribute and topological perturbation. However, these augmentation methods inevitably suffer from the loss of essential information or introduce class-redundant noise. Due to the significant impact of the quality of generated views on contrastive learning, \cite{wu2023sega} theoretically proves that the anchor view containing essential semantic information should have the minimum structural entropy. Inspired by this, structural entropy is employed by us to generate the anchor view with minimum uncertainty.

\mypara{Structural Entropy}.  After Shannon proposed information entropy to measure system uncertainty \cite{shannon1948mathematical}, the measurement of uncertainty in graph structure has been widely studied, and several methods~\cite{rashevsky1955life, trucco1956note, mowshowitz2012entropy} have been developed to quantify it. Among them, structural entropy~\cite{li2016structural,cao2024multi,zeng2023unsupervised,zeng2023hierarchical,zeng2023effective} has been widely used in recent years and has shown promising results in graph structure learning~\cite{zou2023se}, graph pooling\cite{wu2022structural}, and other tasks. Since structural entropy can be used as a metric to measure the complexity of graph hierarchical structure, previous applications have mainly focused on minimizing the structural entropy of the constructed encoding tree. For instance, SEP~\cite{wu2022structural} defines MERGE, REMOVE, and FILL operations to update the constructed encoding tree based on the principle of minimizing structural entropy. In this paper, we employ structural entropy in a self-supervised manner to capture information at both the node level and subgraph level.
\section{PRELIMINARIES}
In this section, we first illustrate the graph-based social bot detection task, followed by an introduction to the definition of graph contrastive learning.

\textbf{Definition 3.1. Graph-based Social bot detection.} Graph-based social bot detection can be regarded as a semi-supervised node binary classification problem on a multi-relational graph. It involves treating the accounts in social platforms as nodes and the interactions such as "following" and "follower" as edges of different relations. Constructed graph in this task can be formulated as $\mathcal{G}=\left\{\mathcal{V},\mathcal{E}, \mathcal{X}\right\}$, where $\mathcal{V}=\{v_{i} \mid i=1,2, \ldots, n\}$ is the set of all nodes, $\mathcal{E}=\cup_{r=1}^R\mathcal{E}_r$ represents the set of edges formed by $R$ relations and $\mathcal{X}$ is the feature matrix. Row $i$ of $\mathcal{X}$ represents the feature vector of the $i$-th node. Total detection process is to use the graph $\mathcal{G}$ and the labels of training nodes $\mathrm{Y}_{\text {train }}$ to predict the labels of test nodes ${\mathrm{Y}}_{\text {test }}$:
\begin{equation}
    f\left(\mathcal{G}, \mathrm{Y}_{\text {train }}\right) \rightarrow{\bar{\mathrm{Y}}}_{\text {test }}.
\end{equation}

\begin{figure*}[t]
    \centering
    \includegraphics[width=0.90\linewidth]{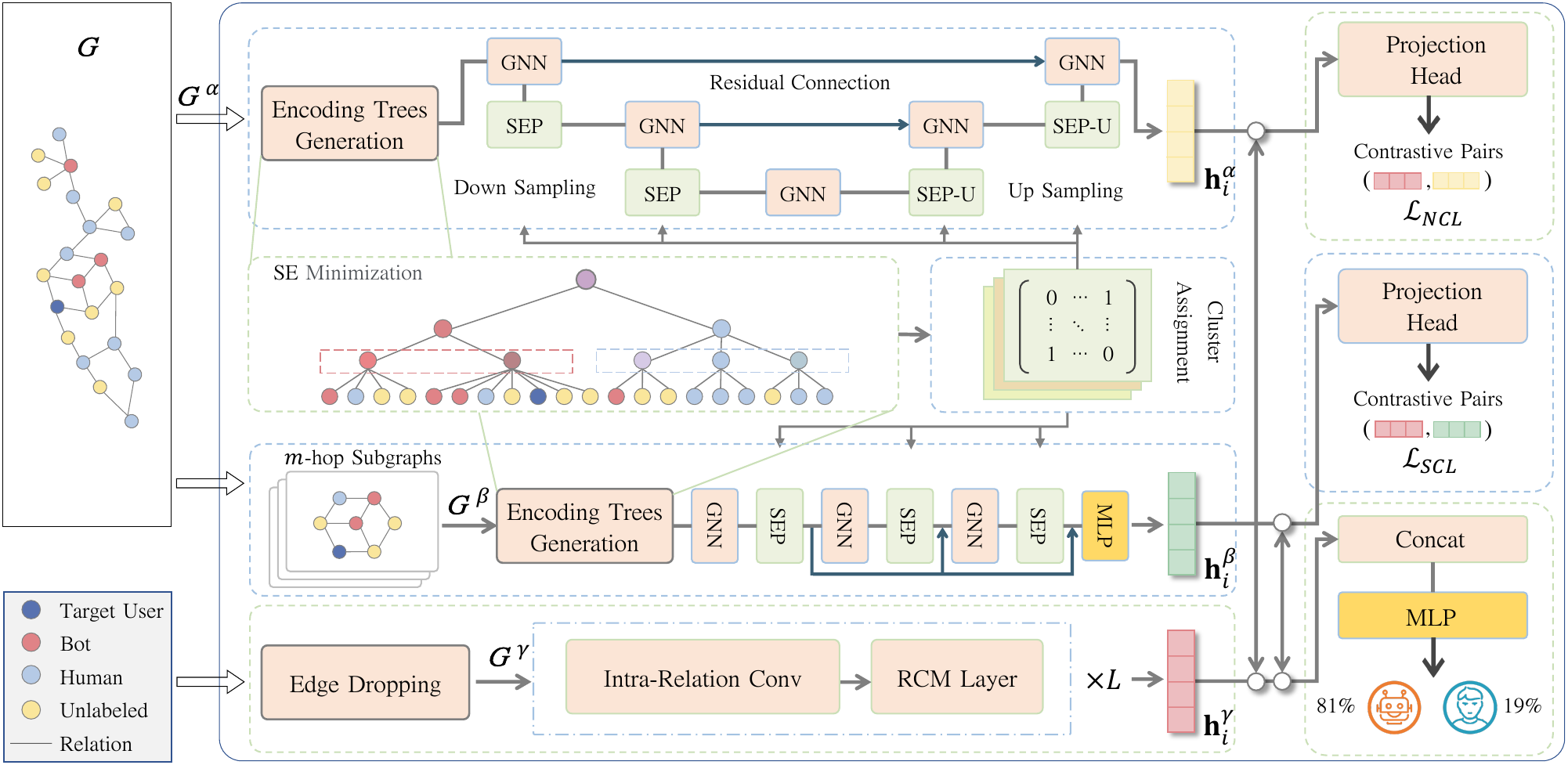}
    \caption{Overview of our proposed framework \SEBot, which mainly consists of three modules: 1) Node-level encoding tree generation and bottom-up message passing; 2) Subgraph-level encoding trees generation and message-passing; 3) Relational information aggregation beyond homophily. Contrastive learning loss and classification loss are later calculated on obtained tree types of representations.} 
    \Description{The total pipeline of \SEBot.}
    \label{fig:framework}
\end{figure*}

\textbf{Definition 3.2. Graph Contrastive Learning.} 
In the general graph contrastive learning paradigm for node classification, two augmented graphs $\mathcal{G}^\alpha$, and $\mathcal{G}^\beta$ are generated using different graph augmentation methods (such as edge dropping, feature masking, etc.) on the input graph $\mathcal{G}$. Subsequently, an encoder consisting of multi-layer graph neural networks is employed to generate node representations including topological information existing in graph structure. During the first training stage, these representations are further mapped into an embedding space by a shared projection head for contrastive learning. A typical graph contrastive loss, InfoNCE~\cite{oord2018representation}, treats the same node in different views $v_i^\alpha$ and $v_i^\beta$   as positive pairs and other nodes as negative pairs. The graph contrastive learning loss $ \mathcal{L}_i $ of node $v_i$ and total loss $\mathcal{L}$ can be formulated as: 
\begin{equation}
\begin{gathered}
     \mathcal{L}_i^\alpha  = -log(\frac{e^{sim(\mathbf{z_i^\alpha},\mathbf{z_i^\beta})/\tau  }}{\sum_{j=1}^{N} \mathbbm{1}_{j \neq i} e^{sim(\mathbf{z_i^\alpha},\mathbf{z_j^\alpha})/\tau  }+ e^{sim(\mathbf{z_i^\alpha},\mathbf{z_j^\beta})/\tau  }} ), \\
     \mathcal{L} = InfoNCE(\mathbf{Z}^\alpha,\mathbf{Z}^\beta )= \sum_{i=1}^N \mathcal{L}_i^\alpha + \mathcal{L}_i^\beta,
\end{gathered}
\end{equation}
where $N$ is the batch size, $\tau$ is the temperature coefficient, $\mathbbm{1}_{j \neq i} = 1$  when $j\neq i$  and $sim(\cdot, \cdot)$ stands for cosine similarity function.

\section{Methodology}
\subsection{Overview of \SEBot} 
The total pipeline of \SEBot is illustrated in Figure \ref{fig:framework}.
To begin with, an attributed multi-relational graph $\mathcal{G}$ is constructed by representing social network interactions as edges. Subsequently, it is fed into three different modules to obtain representations at multi-grained levels under various receptive field scopes $\mathbf{h}_i=[\mathbf{h}_i^{\alpha}\parallel \mathbf{h}_i^{\beta}\parallel \mathbf{h}_i^{\gamma}]$.
The two modules above are responsible for constructing encoding trees by minimizing structural entropy for the entire graph view $\mathcal{G}^\alpha$ and $m$-order subgraph view $\mathcal{G}^\beta$ respectively. $\mathcal{G}^\beta$ is formed by the target node and surrounding neighbor nodes. The view $\mathcal{G}^\gamma$ is generated by edge removing. Then, bottom-up information propagation is performed according to these encoding trees to obtain node embeddings $\mathbf{h}^\alpha_i$ and $\mathbf{h}^\beta_i$. In the third module, to counteract the intentional evasion behaviors of social bots, we devise a graph convolutional layer that can make neighbors similar or discriminative adaptively while maintaining normalization. Simultaneously, a relational channel-wise mixing (RCM) layer is proposed to integrate information from different relations to obtain $\mathbf{h}^\gamma_i$. Finally, the fusion of the three modules involves computing cross-entropy loss for classification and utilizing graph self-supervised contrastive learning loss to capture shared information between three views, enhancing the classification learning process.

\subsection{Community-aware Hierarchical Augment}
\label{sec:community}
In social networks, some accounts may exhibit more pronounced connections with each other due to shared interests, events, and so on, thus forming communities. However, previous detection methods have not effectively leveraged the community structure information within social networks. To reveal the hierarchical structure within the graph, we utilize structural entropy minimization to obtain fixed-height encoding trees, where the child nodes of each node belong to the same community. For the sake of clarity, we first illustrate the definition of structural entropy and its minimization algorithm. 

\textbf{Structural Entropy.} Structural entropy is initially proposed by \cite{li2016structural} to measure the uncertainty of graph structure information. The structural entropy of a given graph $\mathcal{G}=\left\{\mathcal{V},\mathcal{E}, \mathcal{X}\right\}$ on its encoding Tree $T$ is defined as:

\begin{equation}
    \mathcal{H}^T(\mathcal{G})=-\sum\limits_{v_\tau\in T}\frac{g_{v_\tau}}{vol(\mathcal{V})}\log\frac{vol(v_\tau)}{vol(v_\tau^+)},
\end{equation}
where $v_\tau$ is a node in $T$ except for root node and also stands for a subset $\mathcal{V}_\tau \in \mathcal{V}$, $g_{v_\tau}$ is the number of edges connecting nodes in and outside $\mathcal{V}_\tau$, $v_\tau^+$ is the immediate predecessor of of $v_\tau$ and $vol(v_\tau)$, $vol(v_\tau^+)$ and $vol(\mathcal{V})$ are the sum of degrees of nodes in $v_\tau$, $v_\tau^+$ and $\mathcal{V}$, respectively. The structural entropy of graph $\mathcal{G}$ is the entropy of the encoding tree with the minimum structural entropy: $\mathcal{H}(\mathcal{G})=\min_{\forall T}\{\mathcal{H}^T(\mathcal{G})\}$. According to this definition, structural entropy can be used to decode the hierarchical structure of a given graph into an encoding tree as a measurement of community division. In addition, the generated encoding tree $T$ can be seen as the natural multi-grained hierarchical community division result.

\textbf{Minimization Algorithm.} In addition to the optimal encoding tree with the minimum structural entropy, a fixed level of community partitioning is preferred for the specific scenario. Considering this, the $k$-dimension structural entropy of $\mathcal{G}$ is defined as the optimal encoding tree with a fixed height $k$:

\begin{equation}
    \label{equ: minh}
    \mathcal{H}^{(k)}(\mathcal{G})=\min_{\forall T:\mathrm{Height}(T)=k}\{\mathcal{H}^T(\mathcal{G})\}.
\end{equation}

The total process of generation of a encoding tree with a fixed height $k$ can be divided into two steps: 1) construction of the full-height binary encoding tree and 2) compression of the binary encoding tree to height $k$. Given root node $v_r$ of the encoding tree $T$, all original nodes in graph $\mathcal{G}=(\mathcal{V}, \mathcal{E})$ are treated as leaf nodes. We first define two iterative operations on $T$. 
\vspace{\baselineskip}

\textbf{Definition 4.1.} Assuming $v_c^1$ and $v_c^2$ as two children of root node $v_r$, the function $\mathbf{Merge}(v_c^1, v_c^2)$ is defined as adding a new node $v_i$ as the child of $v_r$ and the parent of $v_c^1$ and $v_c^2$:
\begin{equation}
\begin{gathered}
     v_i.children=\{v_c^1,v_c^2\}, \\
     v_r.children=\{v_i\}\cup v_r.children.
\end{gathered}
\end{equation}

\textbf{Definition 4.2.} Given node $v_{\tau}$ and its parent node $v_{\tau}^+$ in $T$, the function $\mathbf{Drop}(v_{\tau})$ is defined as adding the children of $v_{\tau}$ and itself to the child set of $v_{\tau}^+$:
\begin{equation}
    v_\tau^+.children=v_\tau^+.children\cup v_\tau.children.
\end{equation}

The generation of the encoding tree with a fixed height $k$ primarily involves iterations through two operations to obtain the minimum structural entropy, which is shown in Algorithm \ref{algorithm: tree}. To start with, we initialize an encoding tree $T$ by treating all nodes in $\mathcal{V}$ as children of root node $v_r$. During step 1, an iterative $\mathbf{Merge}(v_c^1, v_c^2)$ is conducted with the aim of minimizing structural entropy to obtain a binary coding tree without height limitation. In this way, selected leaf nodes are combined to form new community divisions with minimal structural entropy. Then, to compress height to a specific hyperparameter $k$, $\mathbf{Drop}(v_{\tau})$ is leveraged to merge small divisions into larger ones, and thus the height of the coding tree is reduced, which is still following the structural entropy minimization strategy. Eventually, the encoding tree with fixed height $k$ and minimal structural entropy is obtained. 


\begin{algorithm}[!t]
\caption{Structural entropy minimization algorithm}
\label{algorithm: tree}
\SetKwInOut{Input}{Input}\SetKwInOut{Output}{Output}
\Input{input undirected graph $\mathcal{G}=(\mathcal{V},\mathcal{E})$ and a specific height $k>1$}
\Output{encoding tree $T$ with a fixed height $k$}
    initialize encoding tree $T$ with root node $v_r$ and nodes in $\mathcal{V}$ as its children \;
    // Step 1: full-height binary coding tree construction \;
    \While{$|v_r.children|$ > 2 }{
        select child node pair ($v_c^1$,$v_c^2$) $\leftarrow argmax_{(v_c^1,v_c^2)}\{\mathcal{H}^{T}(\mathcal{G})-\mathcal{H}^{T_c}(\mathcal{G})|v_c^1,v_c^2\in v_r.children\}$ \;
        $\mathbf{Merge}(v_c^1, v_c^2)$\;
    }
    // Step 2: binary coding tree squeeze to height $k$ \;
    \While{$Height(T)$ > $k$}{
        select node $v_\tau$ $\leftarrow$ $argmin_{(v_\tau)}\{\mathcal{H}^{T}(\mathcal{G})-\mathcal{H}^{T_m}(\mathcal{G})|v_{\tau} \in T \& v_{\tau} \neq v_r \& v_{\tau} \notin \mathcal{V}     \}$ \;
        $\mathbf{Drop}(v_{\tau})$ \;
    }
    \Return encoding tree $T$ \;
\end{algorithm}

Obtained encoding tree in this manner can also be seen as the anchor view that includes minimal but sufficient important information in an unsupervised manner~\cite{wu2023sega}. The quality of views generated by graph augmentation technologies plays a crucial role in learning informative representations. According to graph information bottleneck theory (GIB)~\cite{wu2020graph}, retaining important information in a graph view should involve maximizing mutual information between the output and labels while reducing mutual information between input and output. This can be formally expressed as follows:
\begin{equation}
\label{equ: gib}
    \text{GIB}: \max I(f(G);Y)-\beta I(G;f(G)),
\end{equation}
where $I(\cdot;\cdot)$ represents the mutual information between inputs.
A lot of graph augmentation methods have been proposed, including edge removal, feature masking, graph diffusion, and more. Yet, these methods cannot ensure the preservation of essential information regarding downstream tasks and may introduce class-redundant noise. 

\subsection{Message Passing on Encoding Tree}
\label{sec:message}
To obtain node representations and subgraph representations, the message passing on the encoding tree is carried out bottom-up, where the generated parent nodes aggregate information from their child nodes. This process begins with the leaf nodes (i.e., the nodes in $\mathcal{V}$) at the first layer transmitting information to their second-layer parent nodes. Specifically, given the cluster assignment matrix $\mathbf{S}_t\in\mathbb{R}^{n_{t}\times n_{t+1}}$ where $n_{t}$ and $n_{t+1}$ are the number of nodes and assigned clusters (i.e. nodes in the next layer) in the $t$-th layer, each element in $\mathbf{S}_t$ equal to 1 indicates that the node belongs to a corresponding cluster.  The adjacency matrix $\mathbf{A}_{t+1}\in\mathbb{R}^{n_{t+1}\times n_{t+1}}$ and the hidden layer representations $\mathbf{P}_{t+1}\in\mathbb{R}^{n_{t+1}\times d}$ for the $(t+1)$-th layer can be obtained by matrix multiplication:

\begin{equation}
\label{equ: sep}
\textbf{SEP}: \mathbf{A}_{t+1}=\mathbf{S}_t^\top\mathbf{A}_t\mathbf{S}_t;\quad\mathbf{P}_{t+1}=\mathbf{S}_t^\top \mathbf{H}_t,
\end{equation}
where $\mathbf{A}_t$ is the adjacency matrix and $\mathbf{H}_t$ denotes the hidden features martix in the $t$-th layer. As pooling continues, the number of nodes decreases. To obtain representations of nodes in $\mathcal{V}$, we further employ unpooling to ensure that the number of nodes matches the number of nodes in $\mathcal{V}$:

\begin{equation}    
\label{equ: sep-u}
\textbf{SEP-U}: \mathbf{A}_{t+1}=\mathbf{S}_t\mathbf{A}_t\mathbf{S}_t^\top;\quad\mathbf{P}_{t+1}=\mathbf{S}_t\mathbf{H}_t,
\end{equation}
where $\mathbf{S}_t$ is the same matrix used in previous pooling layers. The node-level representation $\mathbf{h}_i^\alpha$ is obtained through multiple layers of SEP to obtain high-order community representations and multi-layer SEP-U reconstructions.

On the other hand, the representation of each subgraph extracted for every target user can be obtained by concatenating the results of SEP pooling layers:

\begin{equation}
\begin{gathered}
    \label{equ: sep-g}
\textbf{SEP-G}:
\mathbf{h}_i^{\beta,t} = Readout(SEP_t(GCN_t(\mathbf{H}_t,\mathbf{A}_t),\mathbf{S}_t)) \; \\
\mathbf{h}_i^{\beta} = Concat(\mathbf{h}_i^{\beta,1}, \cdots, \mathbf{h}_i^{\beta,l}),
\end{gathered}
\end{equation}
where $SEP_t(\cdot)$ denotes the $t$-th SEP layer and $Readout(\cdot)$ can be chosen between average pooling and sum pooling. These high-order representations obtained in this manner enable social bot detection to be implemented through subgraph classification.  
\subsection{Relational Information Aggregation}
\label{sec:relational}
To alleviate graph adversarial attacks by social bots (i.e., actively establishing relationships with humans), we propose a relational information aggregation mechanism beyond homophily and a relation channel-wise mixing layer in this subsection.

\subsubsection{Relational graph convolution beyond resemblance limitation.} Previous work \cite{feng2021botrgcn} has adopted RGCN to detect bot accounts and has shown promising success in modeling different relations. However, the information aggregation of RGCN is based on the homophily assumption (i.e., nodes belonging to the same class tend to be connected), and advanced bots may consciously interact more with humans. Considering this issue, we incorporate high-frequency information (i.e., the differences between nodes) into the information aggregation strategy of RGCN through the generation of negative attention coefficients. However, positive and negative weights generated directly through the tanh activation function can not be normalized. To ensure the consistency of information aggregation, we further introduce the Gumbel-Max reparametrization trick~\cite{jang2016categorical} to make the weights close to 1 or -1. Specifically, the attention weight $\omega_{ij}^{r,\{l\}}$ of the edge $e_{ij}$ in relation $r$ and the layer $l$ is generated by:

\begin{equation}
\label{equ: edge_weight}
\omega_{ij}^{r,\{l\}}=\tanh\left((\mathbf{g}^{\{l\},\top}_r\left[\mathbf{h}_i^{\{l-1\}}\parallel\mathbf{h}^{\{l-1\}}_j\right]+log\epsilon-log(1-\epsilon ) )/\tau_G\right),
\end{equation}
where $\mathbf{g}^{\{l\}}_r\in \mathbb{R}^{2D}$ denotes the trainabele parameter, $\epsilon \sim Uniform(0,1)$ is the sampled Gumbel random variate and $\tau_G$ is a small temperature used to amplify $\omega_{ij}^{r,\{l\}}$. In this manner, when the weights are close to 1, it retains similar information with neighbors, whereas when close to -1, it preserves dissimilar information. In addition, normalization can be directly performed based on the number of neighbors:

\begin{equation}
\label{equ: agg}
    \textbf{h}_i^{r,\{l\}} = \frac{1}{\left | N_{r}(v_i) \right |} \sum_{v_j \in N_{r}(v_i) } \omega_{ij}^{r,\{l\}} \textbf{h}_j^{\{l-1\}}W_r^{\{l\}}, 
\end{equation}
where $N_{r}(v_i)$ is the neighbors of $v_i$ and $W_r^{\{l\}}$ is the weight matrix. Due to the adopted parametrization tick, normalization is approximately satisfied by $| N_{r}(v_i) | \approx \sum_{v_j \in N_{r}(v_i) } |\omega_{ij}^{r,\{l\}}| $.
\subsubsection{Relational Channel-wise Mixing.} After aggregating information independently across different relations, it is necessary to merge this relational information to acquire representations with richer semantics. Previous adaptive method~\cite{feng2022heterogeneity} is limited to generating normalized weight coefficients for representations in each relation, inevitably introducing some noise information present in specific relations. Inspired by~\cite{luan2021heterophily}, we propose a relational channel-wise adaptive fusion layer. Specifically, given the embeddings $ \mathbf{h}_i^{r,\{l\}} $ generated in each relation and at layer $l$, we first calculate the weight vector of the relation:

\begin{equation}
    \label{equ: rcm_1}
    \hat{\mathbf{u }}_i^{r,\{l\}} = [\mathbf{h}_i^{1,\{l\}}\parallel\dots\parallel\mathbf{h}_i^{R,\{l\}}]\hat{W}^{\{l\}}_r,
\end{equation}
where $\hat{W}^{\{l\}}_r$ is the weight matrix and $[\cdot||\cdot]$ is the concat operation.

Then, we apply a channel-wise softmax function to normalize weight coefficients at each feature channel:
\begin{equation}
    \label{equ: rcm_2}
    [\mathbf{u}_i^{1,\{l\}}, \dots, \mathbf{u}_i^{R,\{l\}}]=  Softmax([\hat{\mathbf{u}}_i^{1,\{l\}}, \dots, \hat{\mathbf{u}}_i^{R,\{l\}}]).
\end{equation}

Last, $\mathbf{u }_i^{r,\{l\}}$ can be used to mix representations from different relations and add transformed original features:
\begin{equation}
    \label{equ: rcm_3}
    \mathbf{h}_i^{\{l\}}= \textbf{h}_i^{\{l-1\}}W_{root}^{\{l\}}+\sum_{r=1}^R \mathbf{h}_i^{r, \{l\}} \odot \mathbf{u }_i^{r,\{l\}},
\end{equation}
where $\odot$ denotes the Hadamard product operation and $W_{root}^{\{l\}}$ is the weight matrix to transform original features. 

In fact, when each edge weight $\omega_{ij}^{r,\{l\}}=1$ and $\mathbf{u }_i^{r,\{l\}}=\mathbf{1}$, the proposed encoder is equal to RGCN and thus can be seen as an extension of it. Besides, $L$ layers of the two networks described above are used to generate $\mathbf{h}_i^\gamma$ on $\mathcal{G}^\gamma$.

\subsection{Multi-Task Optimization and Learning}
After obtaining node representations from the three modules, it is necessary to design the loss function as the learning objective. The primary objective of the model remains classification, to identify bot accounts. We employ a two-layer MLP as the classification layer for the concatenation $\mathbf{h}_i=[\mathbf{h}_i^{\alpha}\parallel \mathbf{h}_i^{\beta}\parallel \mathbf{h}_i^{\gamma}]$  and calculate the classification loss using binary cross-entropy:

\begin{equation}
\begin{gathered}
    \label{equ: ce_loss}
    \mathcal{L}_{CE}=-\sum_{v_i\in\mathcal{V}_{t}}[y_{i}log(p_{i})+(1-y_{i})log(1-p_{i})] \\
    p_{i}=softmax(\sigma(\textbf{h}_{i}W_1+b_1)W_2+b_2).
\end{gathered}
\end{equation}

Additionally, as shown in Figure~\ref{fig: intro}, in real-world scenarios, hierarchical community structure and heterophily coexist in social networks. Our goal is to learn unified node representations that can incorporate both types of information simultaneously. However, to our knowledge, there are no methods that capture both types of information simultaneously. Therefore, we use three modules to obtain three representations, $\mathbf{h}_i^{\alpha}$, $\mathbf{h}_i^{\beta}$, and $\mathbf{h}_i^{\gamma}$, each containing different semantic information. Specifically, $\mathbf{h}_i^{\alpha}$ includes global hierarchical structure information, $\mathbf{h}_i^{\beta}$ includes local hierarchical structure information, and $\mathbf{h}_i^{\gamma}$ includes specific category information of the neighborhood.

However, representations obtained through independent modules are often \textit{one-sided}. Therefore, we further use self-supervised contrastive learning to capture the consistency between different representations. Specifically, by maximizing the mutual information between different representations. The proposed two level contrastive learning losses $\mathcal{L}_{NCL}$ and $\mathcal{L}_{SCL}$ are computed based on the representations of all nodes (i.e., $\textbf{H}^\alpha$, $\textbf{H}^\beta$ and $\textbf{H}^\gamma$) across different views:

\begin{equation}
\begin{gathered}
    \label{equ: two_loss}
    \mathcal{L}_{NCL} = InfoNCE(\psi_N(\textbf{H}^\alpha), \psi_N(\textbf{H}^\gamma)), \\
    \mathcal{L}_{SCL} = InfoNCE(\psi_S(\textbf{H}^\beta), \psi_S(\textbf{H}^\gamma)),
\end{gathered}
\end{equation}
where $\psi_N(\cdot )$ and $\psi_S(\cdot )$ are defined as projection heads for node-level and subgraph-level contrastive learning, respectively. In this way, two seemingly contradictory challenges are unified and addressed in a self-supervised manner, rather than being simply solved independently.

Ultimately, the overall loss of the proposed method is calculated by summing the aforementioned three learning losses:
\begin{equation}
    \label{equ: total_loss}
    \mathcal{L} = \mathcal{L}_{CE} + \lambda_1 \mathcal{L}_{NCL} + \lambda_2 \mathcal{L}_{SCL}.
\end{equation}
where $\lambda_1$ and $\lambda_2$ are two hyperparameters ranging from 0 to 1, used to adjust the magnitudes and weights of different losses. 

\subsection{Complexity Analysis}
Given a mutli-relational graph $\mathcal{G}=\left\{\mathcal{V},\mathcal{E}\right\}$, $n=|\mathcal{V}|$ and $m=|\mathcal{E}|$, the runtime complexity of Algorithm~\ref{algorithm: tree} is $O(h_{max}(mlog\;n+n))$, in which $h_{max}$ is the height of coding tree $T$ after the first step. In general, the coding tree $T$ tends to be balanced in the process of structural entropy minimization, thus, $h_{max}$ will be around $log\;n$. Besides, a large-scale social network generally has more edges than nodes, i.e.,  $m\gg n$, thus the runtime of Algorithm 1 almost scales linearly in the number of edges. 

The overall time complexity of the three proposed modules is $O(n+m+h_{max}(m \log n + n))$. Specifically, in Section~\ref{sec:message}, the time complexities of SEP, SEP-U, and SEP-G are all $O(n)$. In Section~\ref{sec:relational}, the proposed relational information aggregation has a time complexity of $O(L(R \cdot n + m))$, where $L$ represents the number of layers and $R$ represents the number of relations.
\begin{algorithm}[!t]
\caption{The training process of \SEBot}
\label{algorithm: sebot}
\SetKwInOut{Input}{Input}\SetKwInOut{Output}{Output}
\Input{input undirected graph $\mathcal{G}=(\mathcal{V},\mathcal{E},\mathcal{X})$ and a specific height $k>1$, hyperparameter $m$, temperature $\tau$ and loss weights $\lambda_1$ and $\lambda_2$}
\Output{node representations \textbf{H}}
    generate undirected graph view $\mathcal{G}^\alpha$ and its $m$-hop subgraph view $\mathcal{G}^\beta$ \;
    generate view $\mathcal{G}^\gamma$ through edge dropping \;
    minimizing structural entropy to construct encoding trees for $\mathcal{G}^\alpha$ and $\mathcal{G}^\beta$ according to Algorithm  \ref{algorithm: tree} \;
    \For{$epoch \leftarrow 1,2,\cdots $}{
        \For{$v_i \in \mathcal{V}$}{
        obtain node embedding $\textbf{h}_i^\alpha$ of $v_i$ from $\mathcal{G}^\alpha$ $\leftarrow$ Equation (\ref{equ: sep}-\ref{equ: sep-u}) \;
        obtain node embedding $\textbf{h}_i^\beta$ of $v_i$ from $\mathcal{G}^\beta$ $\leftarrow$ Equation (\ref{equ: sep-g}) \;   
        obtain node embedding $\textbf{h}_i^\gamma$ of $v_i$ from $\mathcal{G}^\gamma$  $\leftarrow$ Equation (\ref{equ: edge_weight}-\ref{equ: rcm_3}) \;
        }
        classification loss $\mathcal{L}_{CE}$ $\leftarrow$ Equation (\ref{equ: ce_loss}) \;
        contrastive loss $\mathcal{L}_{NCL}$ and $\mathcal{L}_{SCL}$ $\leftarrow$ Equation (\ref{equ: two_loss}) \;
        total loss $\mathcal{L}$ $\leftarrow$ Equation (\ref{equ: total_loss}) \;
        loss backward \;
    }
    \Return the predicted label set for the test nodes $\hat{\mathrm{Y}}_{\text {test }}$.
\end{algorithm}
\section{Experiments}
In this paper, we propose the following research questions for a deep evaluation of the proposed \SEBot framework:
\begin{itemize}[leftmargin=*]
    \item \textbf{RQ1:} How does \SEBot perform compared with other baselines?
    \item \textbf{RQ2:} How does \SEBot benefit from its different modules? 
    \item \textbf{RQ3:} How does \SEBot perform concerning different hyperparameters?
    \item \textbf{RQ4:} Can \SEBot generate more class-discriminative node representations than other baselines? 
    \item \textbf{RQ5:} Can \SEBot effectively address the two challenges mentioned in the introduction?
\end{itemize}

\subsection{Experimental Setup}
\subsubsection{Datasets.} \textbf{TwiBot-20}~\cite{feng2021twibot} and \textbf{MGTAB}~\cite{shi2023mgtab} are adopted to evaluate the performance of \SEBot. \textbf{TwiBot-20} contains 229,580 accounts extracted from Twitter and their following and follower interactions. \textbf{MGTAB} consists of 10,199 accounts and 7 types of relations including follower, friend, mentioned, reply, quoted, reference, and hashtag. Details of two datasets are shown in Table \ref{tab: dataset}, in which \textbf{homo} represents the proportion of edges between nodes of the same class. In addition, we follow the same splits of datasets as~\cite{feng2021twibot} and ~\cite{shi2023mgtab} for training, validating, and testing. 
\subsubsection{Baselines.} We compare \SEBot with previous graph-based social bot detection methods as well as typical GNNs beyond homophily and self-supervised graph contrastive learning methods. All selected baselines are listed below:
\begin{itemize}[leftmargin=*]
    \item \textbf{GCN}~\cite{kipf2016semi} is a typical graph convolution network that can also be seen as a low-pass filter.
    \item \textbf{GAT}~\cite{velivckovic2017graph} lies on an attention-based information aggregation mechanism.
    \item \textbf{GraphSage}~\cite{hamilton2017inductive} is an inductive graph neural network capable of predicting node types that were not seen during the training process.
    \item \textbf{FAGCN}~\cite{bo2021beyond} can adaptively utilize neighbor representations to be similar or diverse by breaking the limitation of low-frequency information aggregation.
    \item \textbf{H2GCN}~\cite{zhu2020beyond} introduces three simple designs: separating self and neighbor embeddings, high-order neighbor information, and concatenating node representations.
    \item \textbf{GPRGNN}~\cite{chien2020adaptive} employs a learnable weight for each order of neighbor information.
    \item \textbf{Alhosseini et al.}~\cite{ali2019detect} is the first to employ graph convolutional neural networks for detecting social bots.
    \item \textbf{FriendBot}~\cite{beskow2020you} utilizes network, content, temporal, and user features obtained from the communication network, and employs machine learning classifiers for classification purposes. 
    \item \textbf{RGT}~\cite{feng2022heterogeneity} constructs heterogeneous views through various relations and proposes the Relation Graph Transformer to obtain node representations.
    \item \textbf{BotRGCN}~\cite{feng2021botrgcn} applies relational graph neural networks to perform social bot detection. \textbf{RGT} and \textbf{BotRGCN} are the current SOTA methods for social bot detection using GNNs.
    \item \textbf{DGI}~\cite{velickovic2019deep} captures consistency between nodes and the entire graph by maximizing local-global mutual information.
    \item \textbf{GRACE}~\cite{zhu2020deep} employs feature masking and edge removal as graph augmentation techniques to generate views. It also uses a discriminator to encourage a uniform distribution.
    \item \textbf{GBT}~\cite{bielak2022graph} calculates the empirical cross-correlation matrix using node representations and computes the loss using the Barlow-Twins loss function.
\end{itemize}

\subsubsection{Hyperparameter Setting}
Hyperparameter settings of our experiments on TwiBot-20 and MGTAB are listed in Table \ref{tab: hyper}. We used the AdamW optimizer to update the parameters. The learning rate is set to 0.01, which is larger compared to baseline models like RGT, resulting in faster convergence. The dropout mechanism is employed to prevent overfitting and maintain high generalization capacity.  The weight of two contrastive loss $\lambda_1$ and $\lambda_2$ are set according to sensitive study results. The tree depth is set to 6, which is a compromise between training time and accuracy. Gumbel-Max reparametrization trick temperature $\tau_G$ is set to 0.01 to amplify edge attention.
\subsubsection{Implementation.} Pytorch \cite{paszke2019pytorch} and Pytorch Geometric \cite{fey2019fast}are leveraged to implement \SEBot and other baselines. All experiments are conducted on a cluster with 8 GeForce RTX 3090 GPUs with 24 GB memory, 16 CPU cores, and 264 GB CPU memory. See our \textbf{code}\footnote{\url{https://github.com/846468230/SEBot}} for more details.

\begin{table}
\caption{Statistics  of TwiBot-20 and MGTAB.}
\centering
\label{tab: dataset}
\scalebox{0.82}{
\begin{tabular}{ccccccc}
\hline
\toprule
\specialrule{0em}{1pt}{1pt}
\textbf{Dataset} & \#\textbf{nodes} & \#\textbf{edges} & \textbf{class} & \#\textbf{class} & \#\textbf{relation} & \textbf{homo} \\ 
\specialrule{0em}{1pt}{1pt}
\hline
\specialrule{0em}{1pt}{1pt}
\multirow{2}{*}{\textbf{TwiBot-20}}     & \multirow{2}{*}{229,580}& \multirow{2}{*}{227,979}    &  human & 5,237 & \multirow{2}{*}{2} &  \multirow{2}{*}{0.53} \\
 & & & bot & 6,589 & & \\
\specialrule{0em}{1pt}{1pt}
\hline
\specialrule{0em}{1pt}{1pt}
\multirow{2}{*}{\textbf{MGTAB}}     & \multirow{2}{*}{10,199}& \multirow{2}{*}{1,700,108}    &  human & 7,451 & \multirow{2}{*}{7} &  \multirow{2}{*}{0.84} \\
 & & & bot & 2,748 & & \\
\bottomrule
\hline
\end{tabular}
}
\end{table}

\begin{table}
\caption{Hyerparameter setting on TwiBot-20 and MGTAB.}
\centering
\label{tab: hyper}
\scalebox{0.8}{
\begin{tabular}{ccc|ccc}
\hline
\toprule
\specialrule{0em}{1pt}{1pt}
\textbf{Parameter} & \textbf{T-20} & \textbf{MGTAB} & \textbf{Parameter} & \textbf{T-20} & \textbf{MGTAB} \\ 
\specialrule{0em}{1pt}{1pt}
\hline
\specialrule{0em}{1pt}{1pt}
optimizer & AdamW & AdamW & hidden dimension & 32 & 32 \\
learning rate & 0.01 & 0.01 & subgraph order $m$ & 2 & 1 \\
dropout & 0.5 & 0.5  & L2 regularization & 3e-3 & 3e-3 \\
loss weight $\lambda_1$ & 0.09 & 0.05 & loss weight $\lambda_2$ & 0.03 & 0.05 \\
tree depth $k$ & 6 & 6  & maximum epochs & 70 & 200 \\
temperture $\tau$ & 0.1 & 0.1 &trick temperature $\tau_G$ & 0.01 & 0.01  \\

\bottomrule
\hline
\end{tabular}
}
\end{table}

\begin{table*}[tb]
\caption{Performance comparison on TwiBot-20 and MGTAB in terms of accuracy, F1-score, recall and precision. The best and second-best results are highlighted with \textbf{bold} and \underline{underline}.'-' indicates that the method is not applicable to MGTAB due to lack of unprocessed raw dataset.}
\centering
\label{tab: rq1}
\scalebox{0.85}{
\begin{tabular}{c|cccc|cccc|c}
\hline
\toprule
\specialrule{0em}{1pt}{1pt}
\multirow{2}{*}{\textbf{METHODs}} &\multicolumn{4}{c}{\textbf{TwiBot-20}} & \multicolumn{4}{|c|}{\textbf{MGTAB}} &\multirow{2}{*}{\textbf{TYPE}}\\
\specialrule{0em}{1pt}{1pt}
\cline{2-5}\cline{6-9}
\specialrule{0em}{1pt}{1pt}
\rule{0pt}{5pt}
 & \textbf{Accuracy}  & \textbf{F1-score}  & \textbf{Recall} & \textbf{Precision} & \textbf{Accuracy}  & \textbf{F1-score}  & \textbf{Recall} & \textbf{Precision} &\\
\specialrule{0em}{1pt}{1pt}
\hline
\specialrule{0em}{1pt}{1pt}
\textbf{GCN}     & 77.53$\pm$1.73 & 80.86$\pm$0.86 & 87.62$\pm$3.31 & 75.23$\pm$3.08 & 80.07$\pm$0.77    &51.71$\pm$4.05    & 75.07$\pm$5.93    & 40.08$\pm$5.90 & \multirow{3}{*}{\textbf{CLASSIC}}\\
\textbf{GAT}      & 83.27$\pm$0.56 & 85.25$\pm$0.38 & 89.53$\pm$0.87 & 81.39$\pm$1.18  & 85.07$\pm$1.19   & 69.32$\pm$4.02    &77.33$\pm$2.19  & 63.34$\pm$7.31 &\\
\textbf{GraphSage} & 85.44$\pm$0.43    & 86.68$\pm$0.61    & 87.66$\pm$2.08    & \textbf{85.78$\pm$0.86}  & 88.58$\pm$1.11    & 77.55$\pm$2.00    & \underline{82.43$\pm$2.39}    & 73.32$\pm$3.12 &\\
\specialrule{0em}{1pt}{1pt}
\hline
\specialrule{0em}{1pt}{1pt}
\textbf{FAGCN} & 85.43$\pm$0.40 & 87.36$\pm$0.32 &  93.00$\pm$0.73 &  82.39$\pm$0.70  
      & 88.11$\pm$1.43    & 77.43$\pm$3.20    & 76.36$\pm$7.90    & 79.19$\pm$3.41 &\multirow{3}{*}{\textbf{HETEROPHILY}}\\
\textbf{H2GCN} &   85.84$\pm$0.34 &  87.57$\pm$0.15 & 92.19$\pm$1.56 & 83.44$\pm$1.32 & 
          89.09$\pm$1.16    & 79.99$\pm$1.53    & 81.00$\pm$5.94    & 79.70$\pm$5.01 &\\
\textbf{GPRGNN} & 86.05$\pm$0.34 & 87.50$\pm$0.30 & 90.25$\pm$0.29 &    84.92$\pm$0.41 
  & 89.07$\pm$1.20    & \underline{80.48$\pm$1.62}    & \textbf{83.54$\pm$1.24}    & 77.67$\pm$2.37 &\\
\specialrule{0em}{1pt}{1pt}
\hline
\specialrule{0em}{1pt}{1pt}
\textbf{Alhosseini et al.} &
   59.88$\pm$0.59 &
  72.07$\pm$0.48 &
  \textbf{95.69$\pm$1.93} &
  57.81$\pm$0.43   & -    & -   & -    & - &\multirow{4}{*}{\textbf{SOTAs}} \\

\textbf{BotRGCN}  &85.75$\pm$0.69 & 87.25$\pm$0.74 & 90.19$\pm$1.72 & 84.52$\pm$0.54 & \underline{89.09$\pm$0.61}    & 79.66$\pm$0.82    & 80.22$\pm$3.07    & 79.39$\pm$4.00\\
 \textbf{FriendBot} & 75.89$\pm$0.47 & 79.97$\pm0.34$ & 88.94$\pm$0.59 & 72.64$\pm$0.52 & - & - & - & - &\\
\textbf{RGT}      & \underline{86.57$\pm$0.42} & \underline{ 88.01$\pm$0.42} & 91.06$\pm$0.80 & \underline{85.15$\pm$0.28} & 
89.00$\pm$1.35   & 79.26$\pm$2.87    & 78.51$\pm$6.25    & \underline{80.48$\pm$2.93} &\\
\specialrule{0em}{1pt}{1pt}
\hline
\specialrule{0em}{1pt}{1pt}
\textbf{DGI}      & 84.93$\pm$0.31 & 87.09$\pm$0.36 & \underline{93.94$\pm$1.13} & 81.17$\pm$0.26 & 87.08$\pm$0.98    & 75.67 $\pm$1.62    & 74.61$\pm$2.58    & 76.81$\pm$1.56 & \multirow{3}{*}{\textbf{CONTRASTIVE}}\\
\textbf{GBT}      & 84.74$\pm$0.92 & 86.87$\pm$0.79 & 93.28$\pm$1.14 & 81.29$\pm$0.92 & 84.68$\pm$0.53    & 70.12$\pm$1.33    & 66.87$\pm$2.30    & 73.80$\pm$1.90 &\\
\textbf{GRACE}      & 84.74$\pm$0.88 & 86.90$\pm$0.84 & 93.56$\pm$1.57 & 81.13$\pm$0.55& 83.16$\pm$1.60    & 66.99$\pm$3.90    &  63.83$\pm$6.00    & 70.77$\pm$2.02 &\\
\specialrule{0em}{1pt}{1pt}
\hline
\specialrule{0em}{1pt}{1pt}
\textbf{\SEBot} & \textbf{87.24$\pm$0.10}    & \textbf{88.74$\pm$0.13}             & 92.97$\pm$1.16    & 84.90$\pm$0.79 & \textbf{90.46$\pm$1.44}    & \textbf{82.12$\pm$2.42}    & 81.73$\pm$2.77    & \textbf{82.52$\pm$2.19}  & \textbf{OURs}\\
\specialrule{0em}{1pt}{1pt}
\bottomrule
\hline
\end{tabular}
}
\end{table*}

\subsection{RQ1: Performance Analysis}
To answer \textbf{RQ1}, we evaluate the performance of \SEBot and 11 other baselines on two social bot detection benchmarks. The experimental results are presented in Table \ref{tab: rq1}, which illustrates that:

\begin{itemize}[leftmargin=*]
    \item \SEBot demonstrates superior performance compared to all other baselines on both datasets in terms of Accuracy and F1-score on both datasets. Furthermore, it achieves relatively high results in terms of Recall and Precision on TwiBot-20 and MGTAB, indicating that \SEBot is better at uncovering social bots and possesses stronger robustness. On the other hand, \SEBot demonstrates the better generalization performance across both datasets, a feat that other methods cannot achieve.
    \item Compared to traditional GNNs (i.e., GCN, GAT, and GraphSage), \SEBot not only considers the community structure but also is conscious of the adversarial structure intentionally constructed by social bots, thus exhibiting stronger detection performance. This also highlights the need for extra fine-grained designs when applying graph neural networks to social bot detection.
    \item Compared to GNNs beyond homophily (i.e., FAGCN, H2GCN, GPRGNN), \SEBot extends further into the social bot detection scenario, specifically on multi-relation directed graphs. In addition, better performance also implies the significant importance of considering adversarial heterophily in social bot detection.
    \item Compared with state-of-the-art graph-based social bot detection methods (i.e., BotRGCN and RGT),  \SEBot achieves the best accuracy and F1-score as it further takes into account the structural semantics present in social networks, which proves to be potent for uncovering deeply concealed bots. Meanwhile, previous graph-based detection methods relied on traditional information aggregation patterns and could not fully capture the structural information within the graph.
    \item Compared with typical self-supervised graph contrastive learning methods (i.e., DGI, GBT, and GRACE), \SEBot retains essential information within the graph by minimizing structural entropy, while other graph augmentation methods may unavoidably introduce noise or lead to the loss of crucial information relevant to downstream tasks. It's important to mention that class imbalance in MGTAB significantly impacts the performance of contrastive learning techniques like DGI, leading to suboptimal results across various metrics.
\end{itemize}

\subsection{RQ2: Ablation Study }
To address \textbf{RQ2}, we conducted ablation experiments as follows. We separately removed the encoding tree of the entire graph, encoding trees of subgraphs and the proposed RCM layer, and evaluated the performance of the residual modules. We also substitute RGCN for the encoder, and adopt different policies of graph augmentation. The results are presented in Table \ref{tab: rq2}. Removing different modules from \SEBot resulted in performance degradation on TwiBot-20 and MGTAB datasets, indicating the pivotal role in overall model effectiveness. This emphasizes the importance of co-considering community structure and heterophilic relations in social bot detection. By comparison, graph augmentation, feature masking, or dropout may disrupt the original feature distribution structure, a critical step in node classification, meanwhile, adding edges may introduce some additional structure noise,  and thus lead to model performance degradation.

\begin{table}[tb]
\caption{Ablation study of \SEBot on TwiBot-20 and MGTAB.}
\centering
\label{tab: rq2}
\scalebox{0.79}{
\begin{tabular}{c|cc|cc}
\hline
\toprule
\specialrule{0em}{1pt}{1pt}
\multirow{2}{*}{\textbf{Settings}}    & \multicolumn{2}{c}{\textbf{TwiBot-20}} &\multicolumn{2}{|c}{\textbf{MGTAB}} \\ 
\specialrule{0em}{1pt}{1pt}
\cline{2-3}\cline{4-5}
\specialrule{0em}{1pt}{1pt}
  & \textbf{Accuracy} & \textbf{F1-score}  & \textbf{Accuracy} & \textbf{F1-score} \\
\specialrule{0em}{1pt}{1pt}
\hline
\specialrule{0em}{1pt}{1pt}
\textbf{Full model}           & \textbf{87.24$\pm$0.10} & \textbf{88.74$\pm$0.13} & \textbf{90.46$\pm$1.44}    & \textbf{82.12$\pm$2.42}\\
\specialrule{0em}{1pt}{1pt}
\hline
\specialrule{0em}{1pt}{1pt}
\textbf{w/o entire graph tree}            & 86.39$\pm$0.30 & 87.84$\pm$0.26 & 89.71$\pm$0.98 & 81.09$\pm$1.41\\
\textbf{w/o subgraph trees}            & 86.45$\pm$0.07 & \underline{88.02$\pm$0.09} & 90.07$\pm$1.04 & 81.79$\pm$1.73\\
\textbf{w/o RCM layer}           & 86.24$\pm$0.49 & 87.69$\pm$0.55 & 89.58$\pm$1.52 & 80.51$\pm$2.63\\
\textbf{RGCN as encoder}          & 86.35$\pm$0.24 & 87.96$\pm$0.26 & \underline{90.32$\pm$1.48} & \underline{81.99$\pm$2.44}\\
\specialrule{0em}{1pt}{1pt}
\hline
\specialrule{0em}{1pt}{1pt}
\textbf{Feature Mask}           & 84.93$\pm$0.67 & 87.00$\pm$0.70 & 90.00$\pm$1.28 & 81.59$\pm$1.85\\
\textbf{Feature Dropping}            & 83.39$\pm$0.61 & 85.53$\pm$0.92  & 89.36$\pm$1.48 & 80.58$\pm$1.27\\
\textbf{Edge Adding}           & \underline{86.52$\pm$0.42} & 87.96$\pm$0.53 & 89.29$\pm$1.39 & 80.60$\pm$2.95\\
\specialrule{0em}{1pt}{1pt}
\bottomrule
\hline
\end{tabular}
}
\end{table}

\begin{figure*}
    \centering
    \includegraphics[width=\textwidth]{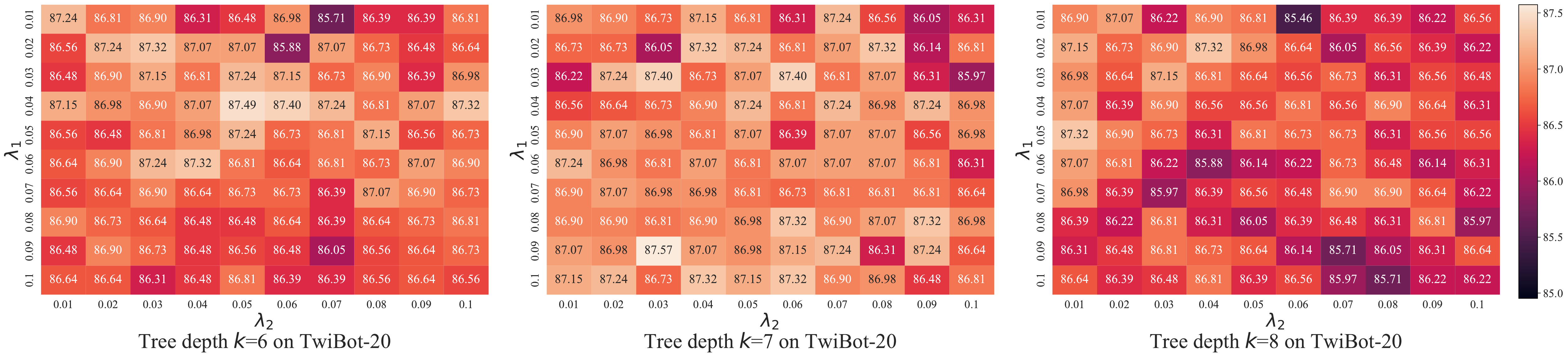}
    \caption{Sensitive analysis of hyperparameter $\lambda_1$ and $\lambda_2$ on TwiBot-20.}
    \label{fig: rq3}
\end{figure*}

\begin{figure*}[]
    \centering
    \subfloat[GCN]{
    \centering
    \includegraphics[width=0.15\textwidth]{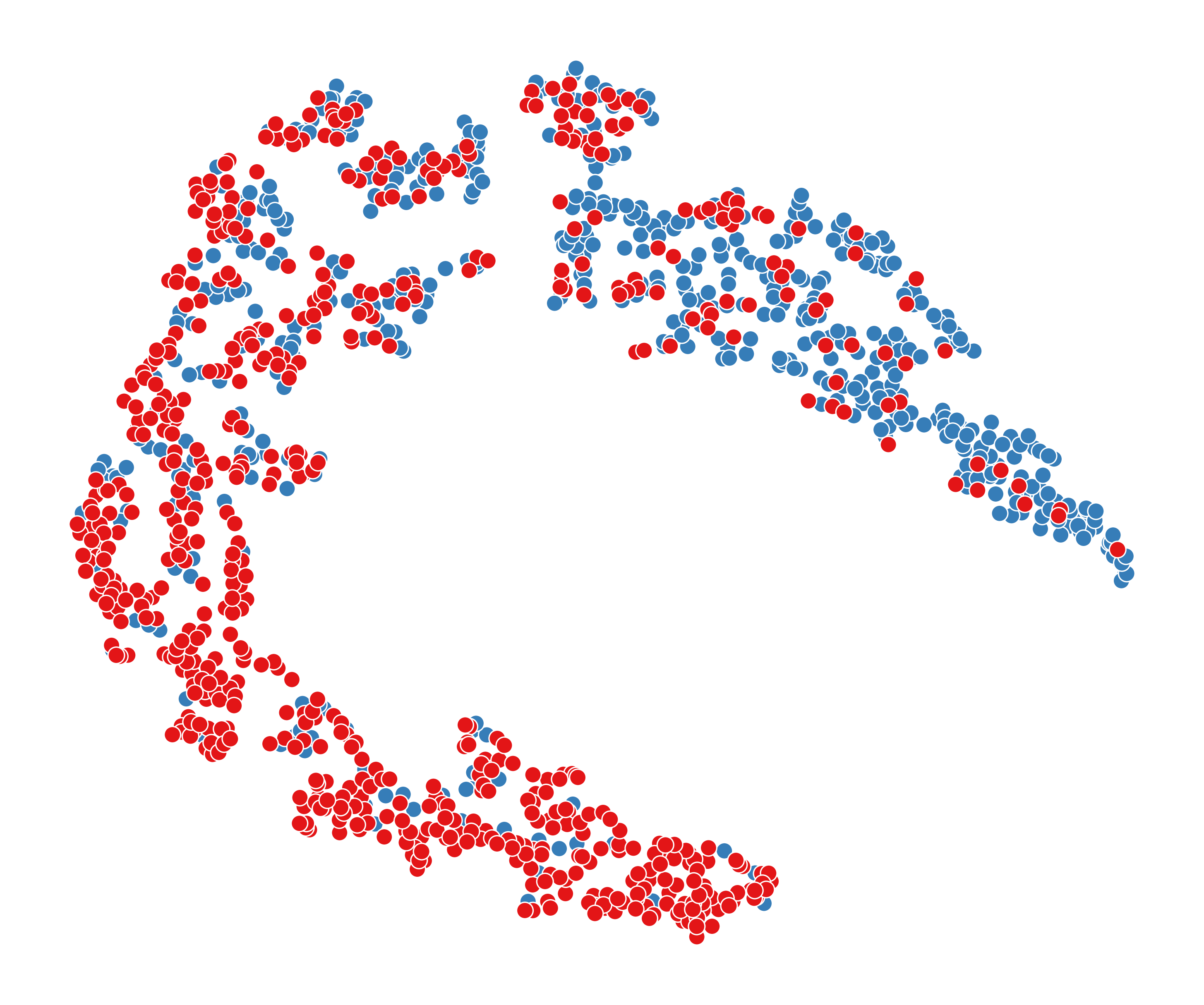}
    }
    \subfloat[FAGCN]{
    \centering
    \includegraphics[width=0.15\textwidth]{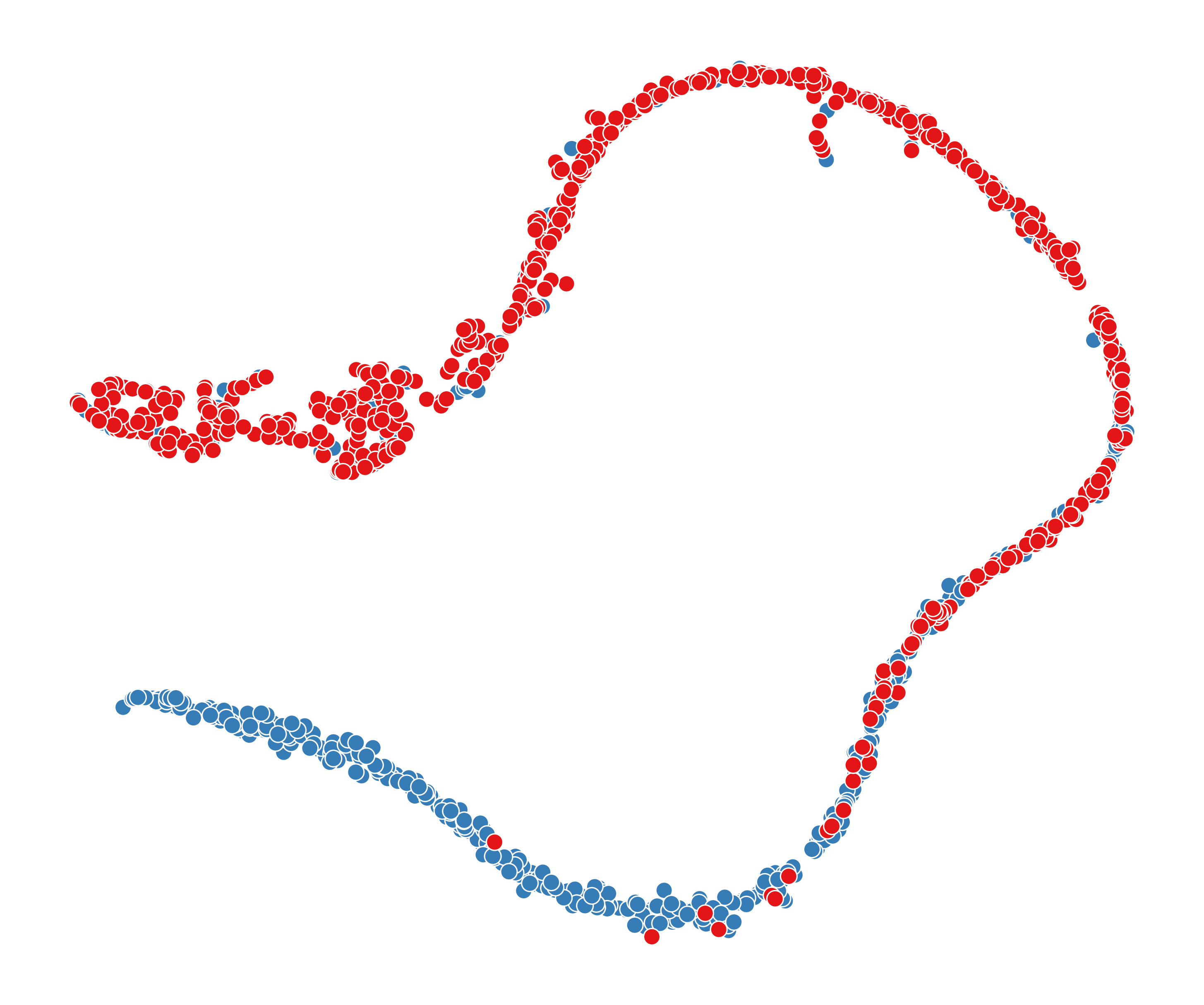}
    }
    \subfloat[BotRGCN]{
    \centering
    \includegraphics[width=0.15\textwidth]{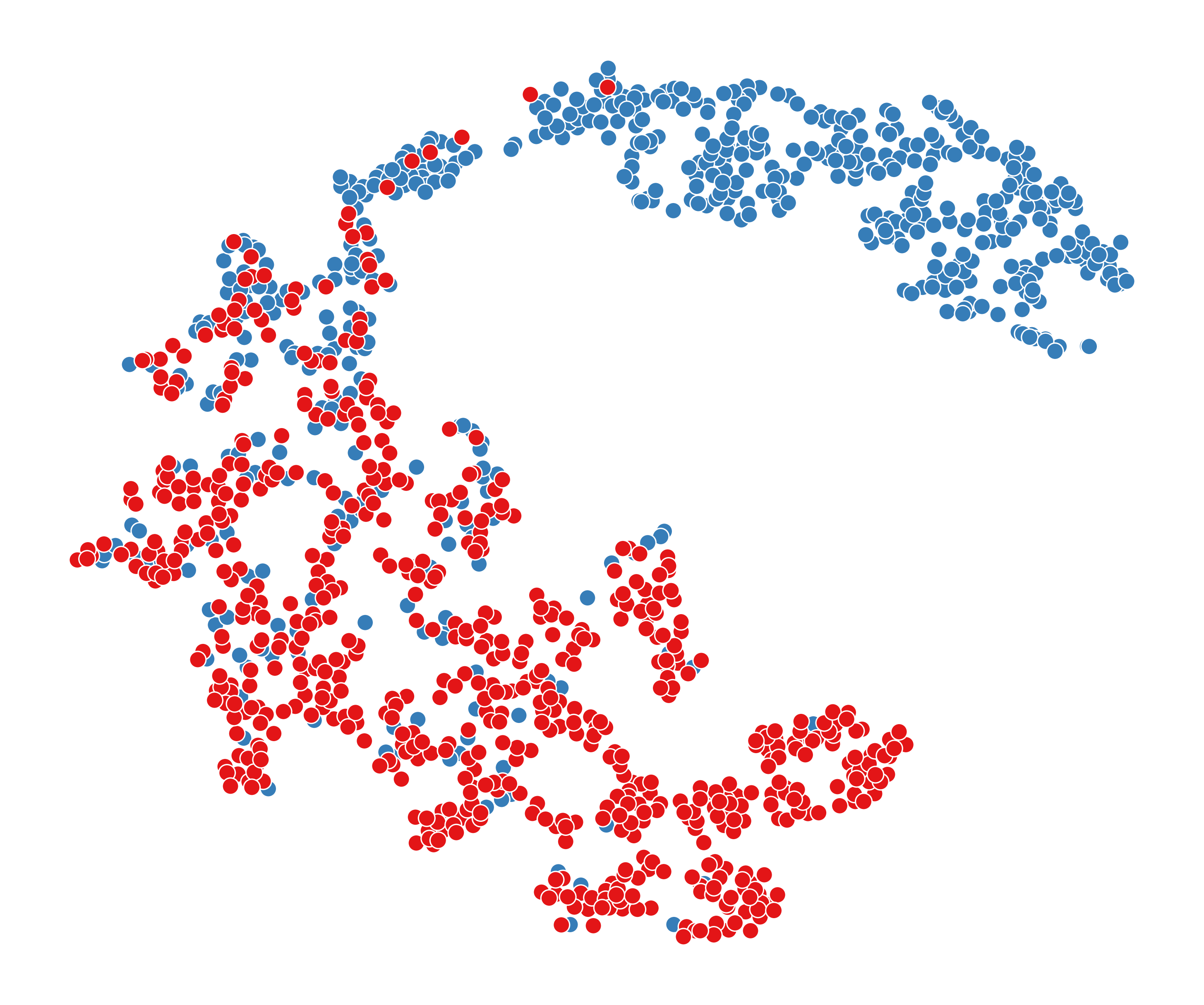}
    }
    \quad
    \subfloat[RGT]{
    \centering
    \includegraphics[width=0.15\textwidth]{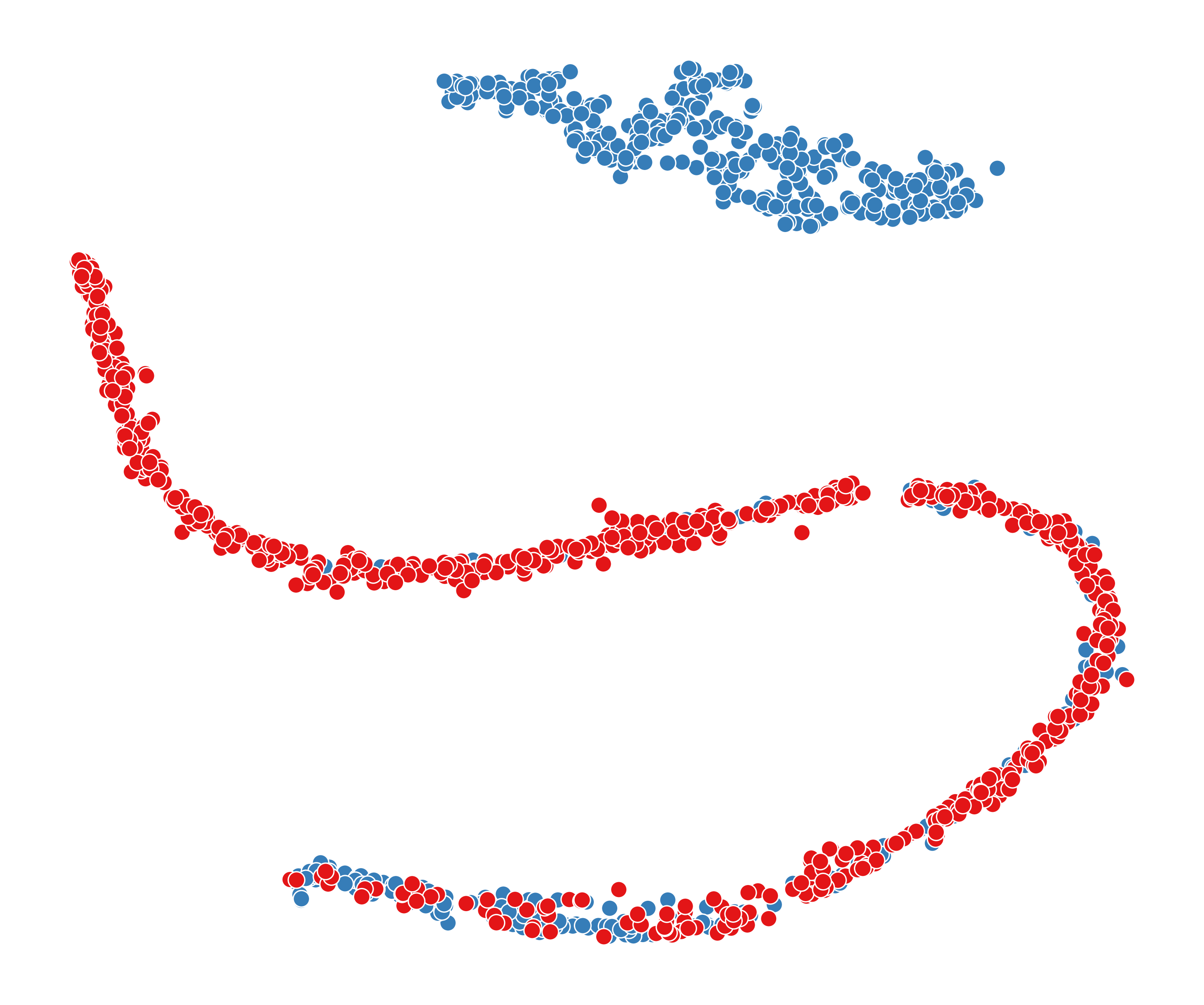}
    }
    \subfloat[GRACE]{
    \centering
    \includegraphics[width=0.15\textwidth]{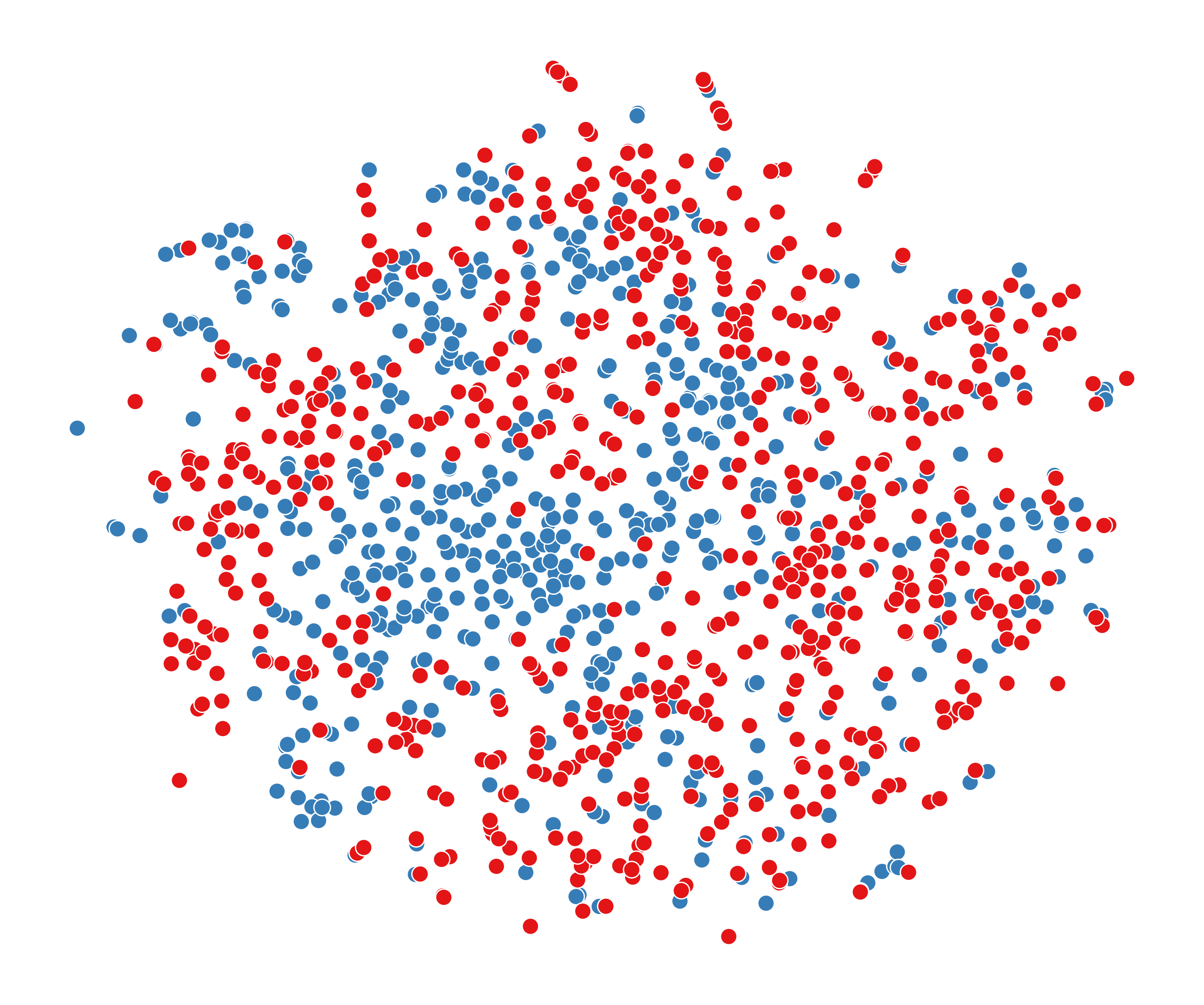}
    }
    \subfloat[\SEBot]{
    \centering
    \includegraphics[width=0.15\textwidth]{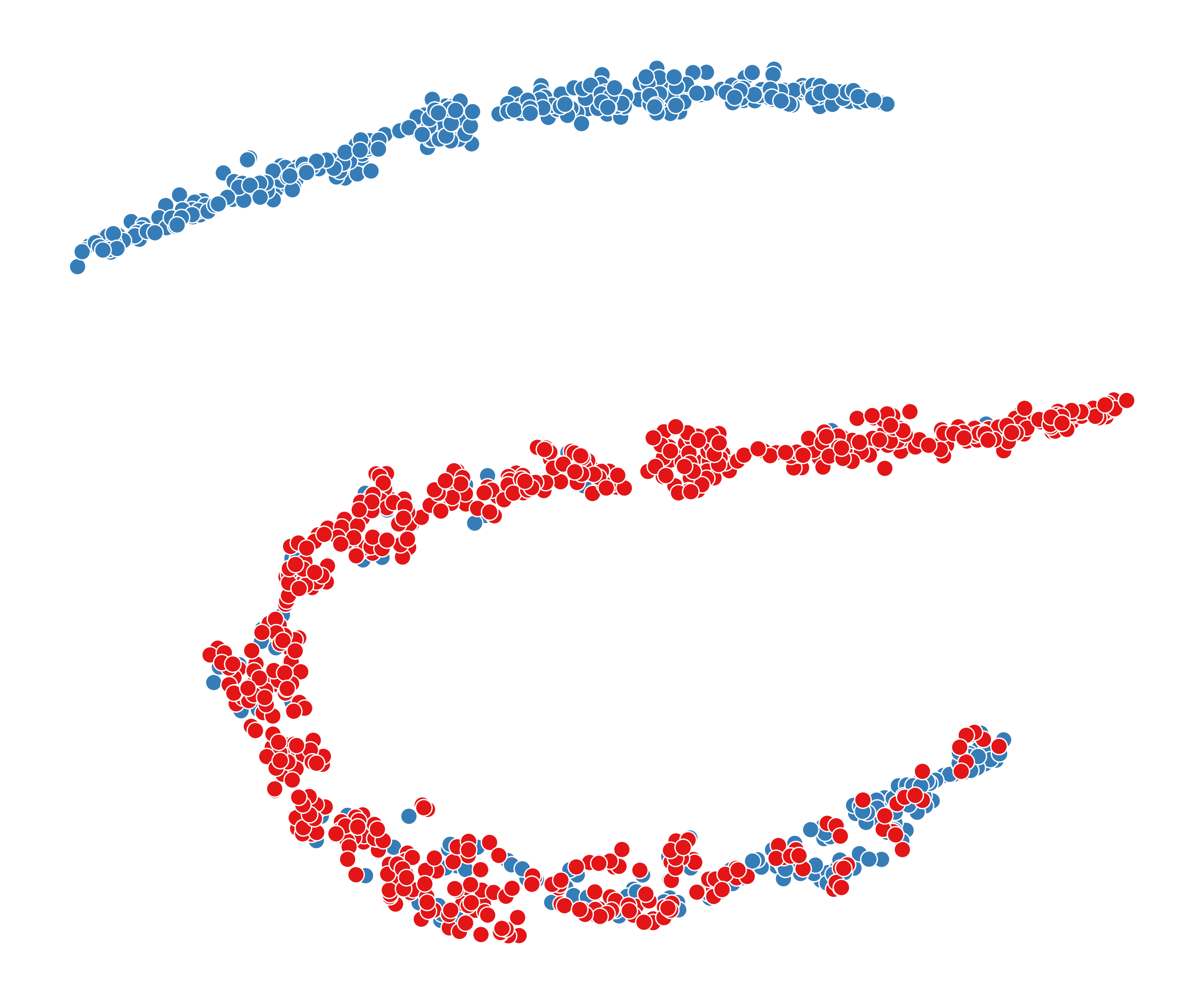}
    }
    \centering
    \caption{Account representations visualization on TwiBot-20. \textcolor[HTML]{E31517}{Red} represents bots, while \textcolor[HTML]{367DB8}{blue} represents humans.}
    \label{fig: rq4}
\end{figure*}
\subsection{RQ3: Sensitive Analysis}
To answer \textbf{RQ3}, we conducted experiments on TwiBot-20 (MGTAB see in Appendix \ref{sec:sas}) to analyze the impact of fixed encoding tree depth and contrastive losses. We set $k$ to 6, 7, and 8, and varied hyperparameters $\lambda_1$ and $\lambda_2$ from 0.01 to 0.1 in steps of 0.01. The results of the experiments are presented in the form of heatmaps in Figure \ref{fig: rq3}. From this, we can intuitively observe that increasing the depth of the tree improves the overall performance of the proposed model. A greater depth enables a finer-grained community partition, benefiting social bot detection, but it requires longer training time. Further, as the tree depth increases, the impact of hyperparameters $\lambda_1$ and $\lambda_2$ on model performance exhibits a fluctuating pattern, initially decreasing and then increasing. Notably, when $k=7$, the model's accuracy shows minimal variation across different hyperparameters. However, increasing or decreasing $k$ results in greater sensitivity of the accuracy to changes in the hyperparameters. This corresponds to our assertion in Section~\ref{sec:community} that, in practical scenarios, a specific tree depth is generally preferred.

\subsection{RQ4: Visualization}
To address \textbf{RQ4}, we visually represent the 128-dimensional node embeddings generated by GCN, FAGCN, BotRGCN, RGT, GRACE, and \SEBot on TwiBot-20 by projecting them onto a 2-dimensional space using T-SNE~\cite{van2008visualizing}, as depicted in Figure \ref{fig: rq4}. The embeddings from GCN and BotRGCN exhibit more scattering, whereas FAGCN, RGT, and our \SEBot produce denser embeddings. GRACE embeddings are quite uniform but suffer from the class collapse issue, where nodes from different classes are located closely to each other~\cite{zheng2021weakly} due to the absence of label information.
It is worth noting that, compared to RGT and FAGCN, our method exhibits local clustering rather than smooth curves visible in 2-dimensional space. This implies that nodes of similar features or belonging to the same community tend to aggregate together in clusters or beads. This also indicates the ability of our method to capture the inherent communities present in the graph structure.
Overall, the embeddings generated by \SEBot demonstrate relatively better class discrimination compared to other methods, and the inclusion of reparameterization techniques ensures that the representations avoid excessive clustering.
\subsection{RQ5: Case Study}
\begin{figure}
    \centering
    \includegraphics[width=0.45\textwidth]{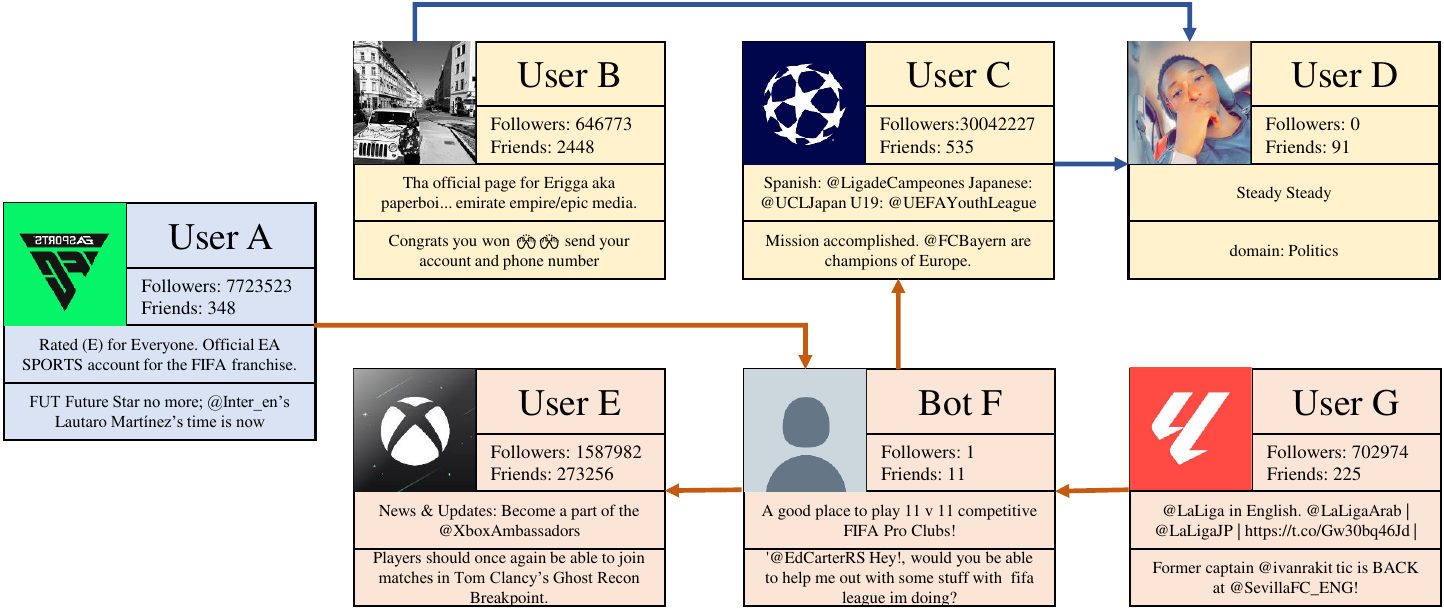}
    \caption{A case study of local community structure and generated edge attention. 
The same background color represents belonging to the same sub-community, with the same parent node on the constructed encoding tree.}
    \label{fig: rq5}
\end{figure}

To answer \textbf{RQ5}, we visualize a local community consisting of 3 subcommunities and 7 social accounts, one of which is a bot account. Due to the employment of the reparameterization technique, edge weights generated adaptively are equal to or near 1 or -1, and edges of two different colors are to represent them, respectively.
As shown in Figure \ref{fig: rq5}, social networks contain both the edges between nodes of the same class and the edges between nodes of different classes. The edge weight adaptive mechanism proposed by us can simultaneously model both types of edges by enabling positive and negative edge weights. 
Furthermore, the fine-grained hierarchical community structure is obtained by minimizing structural entropy with a height constraint. Message passing on the encoding tree provides higher-order feature information favorable for classification. 

\section{Conclusion} 
In this paper, we propose \SEBot, a novel graph-based social bot detection framework that takes into consideration both community structure and adversarial behaviors of social bots. \SEBot addresses the aforementioned issues using three separate modules that leverage structural entropy minimization and a heterophily-aware encoder. \SEBot employs self-supervised contrastive learning to unify and learn the intrinsic characteristics of nodes more effectively. Comprehensive experiments show that \SEBot exhibits superior generalizability and robustness on two real-world datasets compared with all other baselines. 

\begin{acks}
This work was supported by the National Key Research and Development Program of China through the grants 2022YFB3104700, 2022YFB3105405, 2021YFC3300502, NSFC through grants 62322202 and 61932002, Beijing Natural Science Foundation through grant 4222030, Guangdong Basic and Applied Basic Research Foundation through grant 2023B1515120020, Shijiazhuang Science and Technology Plan Project through grant 231130459A.
\end{acks}

\newpage
\bibliographystyle{ACM-Reference-Format}
\bibliography{sample-base}


\clearpage
\appendix
\section{Appendix}
\label{sec:append}

\begin{figure*}[ht]
    \centering
    \subfloat[Sensitive analysis of hyperparameter $\lambda_1$ and $\lambda_2$ on TwiBot-20. (Tree depth 3-5)]{
        \centering
        \includegraphics[width=0.99\textwidth]{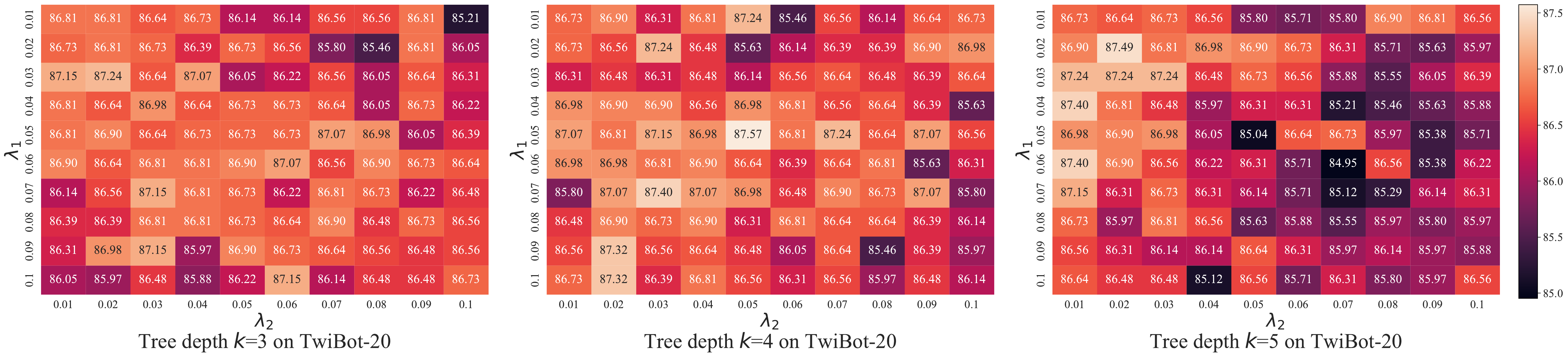}
        \label{fig:rq3-a}
    }
    \quad
    \subfloat[Sensitive analysis of hyperparameter $\lambda_1$ and $\lambda_2$ on MGTAB. (Tree depth 3-5)]{
        \centering
        \includegraphics[width=0.99\textwidth]{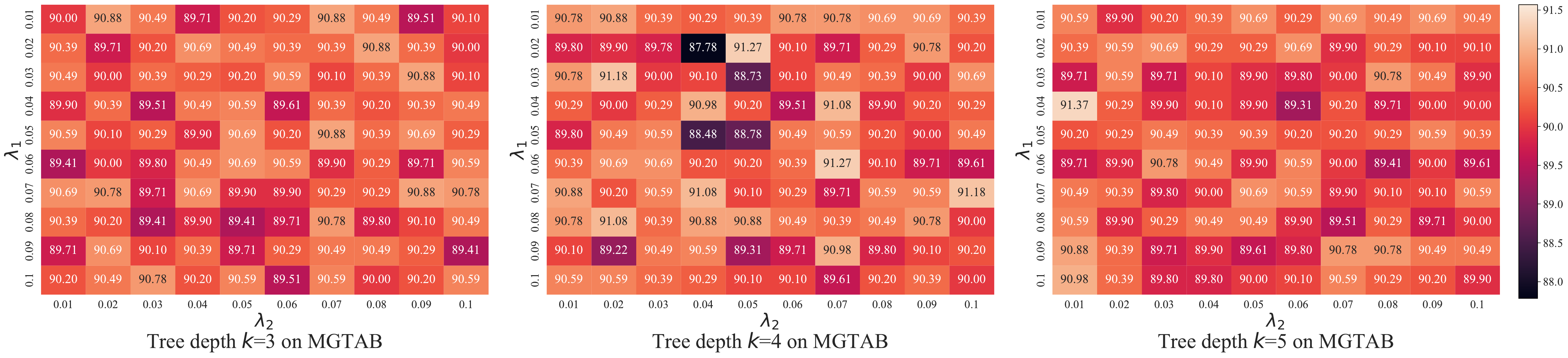}
        \label{fig:rq3-b}
    }
    \quad
    \subfloat[Sensitive analysis of hyperparameter $\lambda_1$ and $\lambda_2$ on MGTAB. (Tree depth 6-8)]{
        \centering
        \includegraphics[width=0.99\textwidth]{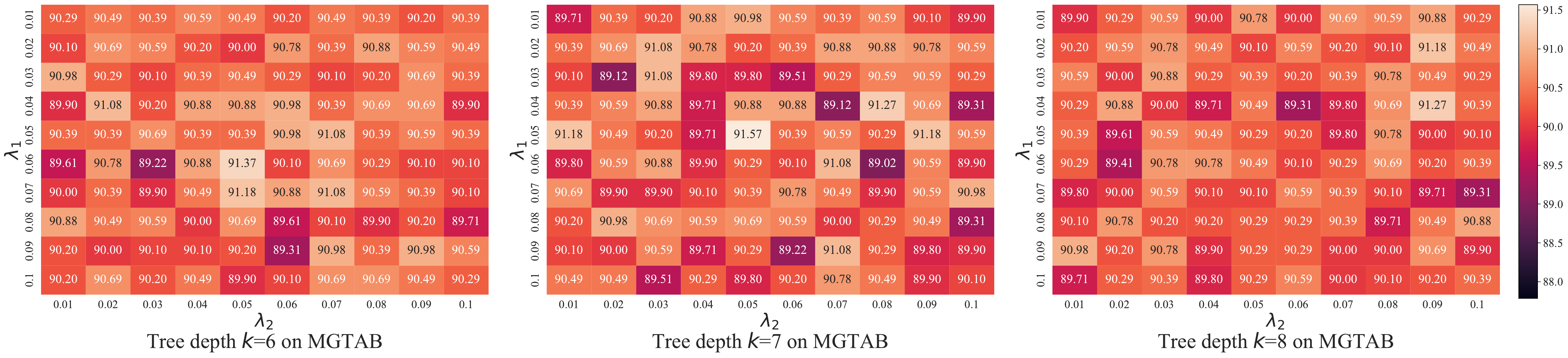}
        \label{fig:rq3-c}
    }
    \caption{Sensitive analysis of hyperparameter $\lambda_1$ and $\lambda_2$ on TwiBot-20 and MGTAB.}
    \label{fig:rq3-3}
\end{figure*}

\begin{figure}
    \centering
    \includegraphics[width=0.9\linewidth]{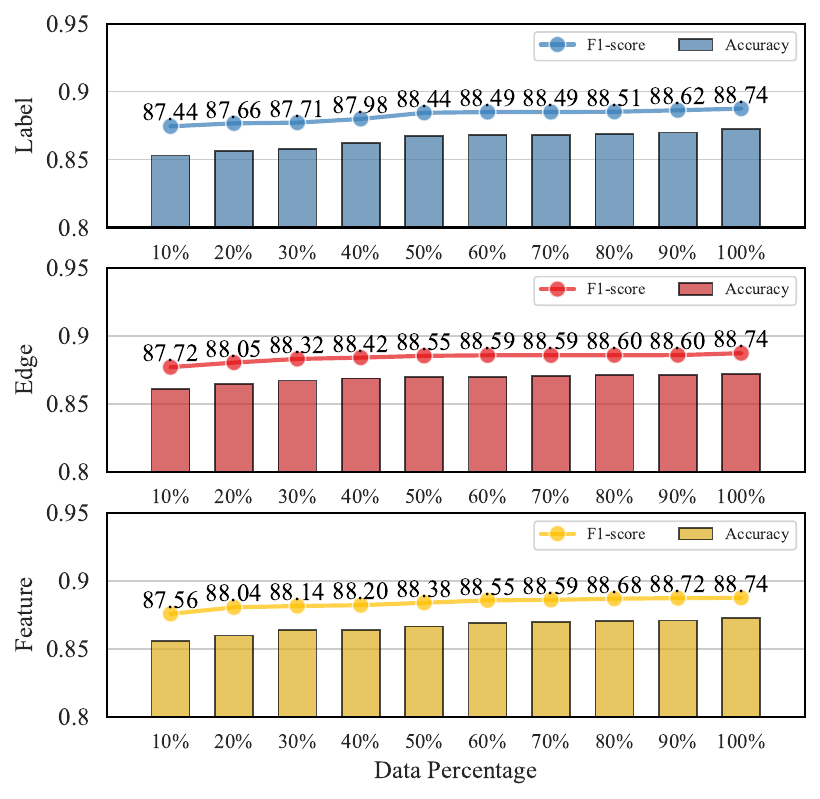}
    \caption{The experimental results of the data efficiency study regarding training data, edge quantity, and feature quantity.}
    \label{fig:exp6}
\end{figure}

\subsection{Sensitive Analysis Supplement}
\label{sec:sas}
In this subsection, we provide additional experimental results on the impact of hyperparameters $\lambda_1$, $\lambda_2$ and tree depth $k$ on accuracy, as shown in Figure \ref{fig:rq3-3}. From Figure \ref{fig:rq3-a}, when $k$ is set to 5, the trained model is more sensitive to changes in hyperparameters $\lambda_1$ and $\lambda_2$, and the impact is larger. When $k$=3, the overall performance is less affected by hyperparameters, and the differences in accuracy are smaller. We also conduct experiments on MGTAB in the same way, which is shown in Figure \ref{fig:rq3-b} and \ref{fig:rq3-c}. It is visually evident that the influence of hyperparameters is relatively small on MGTAB compared to TwiBot-20. Moreover, with the increase in tree depth, there isn't a significant overall improvement in performance. Even when $k$=3, satisfactory results can be achieved, indicating that the performance of \SEBot on MGTAB is minimally affected by the granularity of community partitioning.

\subsection{Data Efficiency Study}
Current approaches to social bot detection predominantly follow a supervised paradigm, heavily reliant on an adequately rich set of annotated training data. However, acquiring such a dataset is a costly endeavor, and issues such as inaccurate annotations, noise, and insufficient richness are widespread. Consequently, it becomes imperative to evaluate the performance of \SEBot under conditions of limited training data, relationships, and account features. To address this, we specifically design experimental conditions by training solely on a subset of the data, randomly removing edges, and masking partial features. The results are illustrated in Figure \ref{fig:exp6}. Notably, even under the constraint of utilizing only 50\% of the training data, \SEBot demonstrates superior performance compared to RGT~\cite{feng2022heterogeneity}, affirming its capability to mitigate dependence on data. Furthermore, our observations indicate that increased edges and features contribute to enhanced performance, suggesting that both factors play an advantageous role in detection." 


\end{document}